\documentclass[twocolumn]{aastex63}

\usepackage{lmodern}
\usepackage{latexsym}
\usepackage{amsmath}
\usepackage{amssymb}
\usepackage{amsbsy}
\usepackage{amsthm}
\usepackage{amsfonts}
\usepackage{mathrsfs}
\usepackage{bm}
\usepackage{sansmath}
\usepackage{relsize}
\usepackage{caption2}
\usepackage{graphicx}
\usepackage[utf8]{inputenc} % usually not needed (loaded by default)
\usepackage[T1]{fontenc}
\usepackage{epstopdf}

\submitjournal{ApJ}

\shorttitle{Wind from hot accretion flow}
\shortauthors{Mosallanezhad et al.}
\begin{document}

\title{Two-dimensional Inflow-Wind Solution of Hot Accretion Flow. I. Hydrodynamics}

\correspondingauthor{Fatemeh Zahra Zeraatgari}
\email{fzeraatgari@xjtu.edu.cn}

\author[0000-0002-4601-7073]{Amin Mosallanezhad}
\affiliation{School of Mathematics and Statistics, Xi'an Jiaotong University, Xi'an, Shaanxi 710049, People's Republic of China}

\author[0000-0003-3345-727X]{Fatemeh Zahra Zeraatgari}
\affiliation{School of Mathematics and Statistics, Xi'an Jiaotong University, Xi'an, Shaanxi 710049, People's Republic of China}

\author[0000-0003-3468-8803]{Liquan Mei}
\affiliation{School of Mathematics and Statistics, Xi'an Jiaotong University, Xi'an, Shaanxi 710049, People's Republic of China}

\author[0000-0002-0427-520X]{De-Fu Bu}

\affiliation{Key Laboratory for Research in Galaxies and Cosmology, Shanghai Astronomical Observatory, Chinese Academy of Sciences, 80 Nandan Road, Shanghai 200030, People's Republic of China}



\begin{abstract}

We solve the two-dimensional hydrodynamic equations of hot accretion flow
in the presence of the thermal conduction. The flow is assumed to be in steady-state 
and axisymmetric, and self-similar approximation is adopted in the radial direction. 
In this hydrodynamic study, we consider the viscous stress tensor to mimic 
the effects of the magnetorotational instability for driving angular momentum.
We impose the physical boundary conditions at both 
the rotation axis and the equatorial plane and obtain the solutions 
in the full $ r-\theta $ space. We have found that thermal conduction 
is indispensable term for investigating the inflow-wind structure of the hot accretion flows 
with very low mass accretion rates.
One of the most interesting results here is that the disc is convectively stable in 
hot accretion mode and in the presence of the thermal conduction. Furthermore, 
the properties of wind and also its driving mechanisms are studied. Our analytical 
results are consistent with previous numerical simulations of hot accretion flow.

\end{abstract}



\section{Introduction} \label{sec:intro}
Various black hole accretion models have been proposed in the past several decades
including standard thin disc (\citealt{Shakura and Sunyaev 1973}), super-Eddington accretion flow
(slim disc, \citealt{Abramowicz et al. 1988}), and also hot accretion flow (\citealt{Narayan and Yi 1994}; 
\citealt{Abramowicz et al. 1995}). 
Based upon the temperature of the accretion flow, these models can be divided into
cold and hot modes where the standard thin disc and super-Eddington accretion flow belong 
to cold mode (\citealt{Yuan and Narayan 2014}).

In recent years, wind has become a fascinating subject in the study of accretion 
flows in both cold and hot modes. Wind appears to carry significant mass, 
angular momentum and energy away from the disc, and has 
the potential for a greater impact on its surrounding. This discernible effect 
on its environment is persuasive enough for interest in this topic.
Further, the elimination of mass and angular momentum from the disc might 
essentially alter the accretion process (\citealt{Shields et al. 1986}). There have been 
substantial observational evidence of wind in cold accretion mode via blue 
shifted absorption lines, in luminous AGNs (\citealt{Crenshaw et al. 2003}; \citealt{Tombesi et al. 2010,Tombesi et al. 2014}; 
\citealt{Liu et al. 2013}; \citealt{King and Pounds 2015}) and X-ray binaries in 
high/soft state (\citealt{Neilsen and Homan 2012}; \citealt{Homan et al. 2016}; \citealt{Diaz Trigo and Boirin 2016}).
Nevertheless, the challenging objects for detection are systems
in hot accretion mode, since the accreting gas is virially hot and fully ionized 
in low-luminous AGNs (LLAGNs) (\citealt{Tombesi et al. 2010,Tombesi et al. 2014}; 
\citealt{Crenshaw and Kraemer 2012}; \citealt{Cheung et al. 2016}) and low/hard state 
of X-ray binaries (\citealt{Homan et al. 2016}; \citealt{Munoz-Darias et al. 2019}). 

Wind can be driven by different physical mechanisms such as thermal, 
radiation and magnetic pressures. 
Radiation pressure as well as magnetic forces would act on smaller scales 
(\citealt{Proga et al. 2000}; \citealt{Fukumura et al. 2015}), while thermal driving might work in outer 
regions of the disc, and could adjust the mass accretion rate through the disc (\citealt{Shakura 
and Sunyaev 1973}; \citealt{Begelman et al. 1983}; \citealt{Shields et al. 1986}). A large number of 
numerical simulations have been done to show the wind is potentially able to transfer 
a significant amount of power from the black hole accreting system in hot accretion mode
(e.g., \citealt{Stone et al. 1999}; \citealt{Igumenshchev and Abramowicz 1999, Igumenshchev and Abramowicz 2000}; 
\citealt{Hawley and Balbus 2002}; \citealt{Pang et al. 2011}; \citealt{Yuan et al. 2012a, Yuan et al. 2012b}; \citealt{Bu et al. 2013}; 
\citealt{Narayan et al. 2012}; \citealt{Li et al. 2013}); however, the actual mechanism 
driving such winds is still a source of much debate. 

Analytical method is often invoked to investigate the existence of wind from accretion 
flow in hydrodynamic (HD) and magnetohydrodynamic (MHD) studies (e.g., \citealt{Bu et al. 2009}; \citealt{Mosallanezhad et al. 2014}; \citealt{Samadi et al. 2017}; \citealt{Bu and Mosallanezhad 2018}; \citealt{Kumar and Gu 2018}; \citealt{Zeraatgari et al. 2020}) 
and calculate the physical properties of the wind. In principle, it is very difficult and 
time consuming to perform numerical simulations with the most updated physical terms as well as
different input parameters to examine the dependency of the results to the initial set
of parameters. Therefore, analytical studies are very powerful tools
for better understanding the dependency of inflow-wind structure of the system 
to the physical parameters and attempt to find the real mechanism of producing wind in
such accreting systems.

The analytical study of hot accretion flow started from height-averaged, 
radially self-similar solutions presented by \citealt{Narayan and Yi 1994} which could not 
show a clear picture of the vertical structure of such a system. Next, in \citealt{Narayan and Yi 1995} 
(henceforth NY95), they revisited their solutions in $ r-\theta $ plane in spherical 
coordinates. They adopted self-similar solutions in radial direction and to solve the 
equations in the $ \theta $-direction used proper boundary conditions at both equatorial 
plane and rotation axis. Unfortunately, the solutions did not 
show an inflow-wind structure, since they considered $ v_{\theta} = 0 $ 
and zero radial velocity at the pole, i.e., $ v_{r}(0) = 0 $.
By eliminating $ v_{\theta} $ and adopting axisymmetric and steady state assumptions, 
the radial self-similar solution of the density followed by: $ \rho \propto r^{-3/2} $ 
(see the first term in equation (\ref{pde_continuity})).   
Their solutions also became singular when $ \gamma = 5/3 $  
albeit the numerical simulations of hot accretion flow suggest that in the non-relativistic cases, 
$ \gamma $ is very close to $ 5/3 $ (e.g. \citealt{Balbus and Hawley 1998}; \citealt{Blandford and Begelman 1999}). 
However, based on their positive value of Bernoulli parameter, they 
argued that bipolar outflow must exist near the rotation axis.

After that, \citealt{Xu and Chen 1997} (henceforth XC97) brought up all components of the velocity 
including $ v_{\theta} $, adopted Fourier series and obtained accretion and ejection solutions.
\citealt{Blandford and Begelman 2004} also represented self-similar two-dimensional solutions 
for radiatively inefficient accretion flows with outflow called adiabatic inflow-outflow 
solutions (ADIOS). They suggest that the mass accretion rate decreases inward
due to the mass loss in the outflow. Based on their solution, the mass inflow and outflow
fluxes follow the power law of $ \dot{M} \propto r^{s} $ with equal and opposite values.
They also determined the power index $ s $ to be in the range $ 0  \leqslant s \leqslant 1 $.

\citealt{Tanaka and Menou 2006} (henceforth TM06) followed NY95 and they mainly focused on 
the effects of the thermal conduction on the global properties of hot accretion flows. 
Even though they did not include $ v_{\theta} $, their solutions obtained positive radial velocity
near the rotation axis interpreting the outflow. Here, we intend to emphasize the requisite role
of $ v_{\theta} $ in wind studies. As an example, XC97 considered $ v_{\theta} $ but some researchers 
argued that the solution would not be correct (e.g., \citealt{Xue and Wang 2005}, \citealt{Jiao and Wu 2011}). 
This is mainly because, they believed the mass flux of the outflow became exactly 
equal to the mass flux of the inflow at a certain radius. 
To avoid this knotty issue, for instance, \citealt{Xue and Wang 2005} truncated the solutions where $ v_r = 0 $
by prescribing the opening angle $ \theta_{0} $ and considered this place as the surface of the disk. 
They set the sound speed on the surface as $ c_{s} / \Omega_{k} \lesssim (\pi/2 - \theta_0)r $. Therefore, the second boundary is set as an input parameter rather than being calculated in their solution. Actually, their solutions were limited to the inflow region near the equatorial plane and a surface from which the wind would blow out.
Also, \citealt{Jiao and Wu 2011} integrated the equations from equatorial plane toward the rotation axis 
and stopped integration where the density or gas pressure became negative.
Although an outflow structure could be shown, their solutions did not reach to the pole, 
$ \theta = 0 $, where some of the physical boundary conditions must be satisfied.

\citealt{Khajenabi and Shadmehri 2013} (henceforth KS13) also solved the HD equations of hot accretion flow in the presence of the thermal conduction as well as all three components of the velocity. Further, they imposed the same physical constraint as \citealt{Xue and Wang 2005} for the opening angle and obtained this angle self-consistently from their numerical integration starting from equatorial plane. Their results showed that the thermal conduction affected the opening angle of the wind as an increase of that would shrink the size of the wind region.

The main aim of this paper is to find inflow-wind solution of hot accretion flow in the whole vertical direction by imposing proper boundary conditions at both the rotation axis and the equatorial plane and then compare the results with those above mentioned studies. In the following section, 
we will find the origin of discrepancy between all those aforementioned analytical 
solutions by focusing on the continuity and the energy equations.
We will illustrate that it would be possible to solve the hydrodynamic equations of hot accretion flow in 
the whole $ \theta $-direction using two point boundary value problem.
  
In this paper, to mimic 
the effects of magnetorotational instability (MRI) in driving angular momentum, we will consider 
the viscous stress tensor. Another aim of the present paper is 
to find the dependency of the results to the initial set of 
parameters such as the conductivity coefficient, density index, and advection parameter.
Note here that, the gradient of the magnetic pressure is one of the 
driving mechanisms for producing wind and plays an important role in the dynamics of hot accretion flows.
In the sense of on this point, we intend to embark on a series of papers to solve hydrodynamic (HD) 
and magnetohydrodynamic (MHD)\footnote{The MHD equations 
will be solved in the second paper of these series and the 
HD and MHD results will be compared there.} equations of hot accretion flow
using appropriate boundary conditions
at the rotation axis as well as the equatorial plane.

The remainder of the paper is organized as follows. In section \ref{sec:answer_to_question}, 
we will investigate the origin of discrepancy between all previous
analytical solutions and find the reason why their solutions could not reach the rotation axis.
The basic HD equations, physical assumptions, self-similar solutions, and also the boundary 
conditions will be introduced in Section \ref{sec:basic_equations}. 
In Section \ref{sec:results}, the detailed explanations of numerical results will be presented.  
Finally, in section \ref{sec:summary_discussion}, we will provide the summary and discussion.
\section{The origin of discrepancy between previous analytical solutions}  \label{sec:answer_to_question}
 
Based on Reynolds transport theorem, the time rate of change of integrals of
physical quantities within material volumes can be calculated as,

\begin{equation}
	\frac{d \mathcal{F}}{d t} = \int_{V} \left( \frac{d F}{d t} + F \nabla \cdot \bm{v} \right) dV,
\end{equation} 
where, $ F(\bm{x},t) $ can be any scalar, vector, or tensor field, with 
$ \mathcal{F} = \int_{V} F(\bm{x},t) dV $ (see equation (18.2) in 
\citealt{Mihalas and Mihalas 1984} for more details). According to the conservation 
law of mass for the fluid, the mass within a material volume must be 
always the same, i.e.,

\begin{equation} 
	\frac{d}{d t} \int_{V} \rho\, dV = 0.
\end{equation}

By applying Reynolds transport theorem for $ F = \rho $, we have:

\begin{equation} \label{Reybolds}
	\int_{V} \left( \frac{ \mathrm{d} \rho}{\mathrm{d} t} +\rho \nabla \cdot \bm{v} \right)\, dV = 0.
\end{equation}

In principle, this integral will be vanished only if the integrand vanishes at all
points in the flow field. The integrand then is the continuity equation (see equation \ref{eq:continuity}).
In spherical coordinates, by imposing the axisymmetric $ (\partial/ \partial \phi = 0) $ 
and steady state $ (\partial /\partial t = 0) $ assumptions into the equation of continuity,
it will be reduced to the equation (\ref{pde_continuity}). Integrating the first term of this equation over angle will give us,

\begin{equation} 
	\dot{M} = - \int 2 \pi  r^{2} \sin \theta \rho v_{r} \, d \theta,
\end{equation}
where, $ \dot{M} $ is the net mass accretion rate. 
In NY95 since $ v_{\theta} = 0 $, the second term was spontaneously 
eliminated from the equation (\ref{pde_continuity}). As a result, the 
density index of their self-similar solutions became $ n = 3/2 $ in 
$  \rho \propto r^{-n} $ (see equations (2.11)-(2.15) of NY95). 
In the case of $ v_{\theta} \neq 0 $, the sum of two terms in the equation (\ref{pde_continuity}) 
must be always zero at each specific radius to satisfy the continuity equation. 
Keeping the second term of equation (\ref{pde_continuity}) would also cause flattening 
of the density profile, i.e., density index would be in the range of $ 0 < n < 3/2 $. 

In XC97, they included $ v_{\theta} $ in their solutions and found 
two different kinds of solutions, i.e., accretion-outflow and ejection-outflow 
for the density index of $ n < 3/2 $ and $ n > 3/2 $, respectively. However, the ejection-outflow with 
the large value for density index of $ n > 3/2 $ might not be correct. The reason would 
emanate from the energy equation which is defined as,
\begin{equation} \label{intro_enegry}
	Q_\mathrm{adv} = Q_{+} - Q_{-}
\end{equation}
where $ Q_\mathrm{adv} $, $ Q_{+} $, and $ Q_{-} $ are energy advection, 
total heating and cooling terms per unit volume, respectively. In hot accretion 
flows or radiative inefficient accretion flows (RIAFs), it is common to omit 
$ Q_{-} $ and introduce advection parameter as $ f \equiv Q_\mathrm{adv} / Q_{+} $ with 
$ 0 < f \le 1 $. Hence, the energy equation of the hot accretion flow in XC97 was written as, 
\begin{equation} \label{energy_Xu_Chen}
	Q_\mathrm{adv} \equiv f Q_{+}
\end{equation}
with the total energy heating released by the viscosity. 
By assuming axisymmetric, steady state, and radially self-similar approximation, 
the advection and the viscous terms of the energy equation were reduced to the 
dimensionless forms as described in equation (\ref{q_adv_XC97}) and 
equation (\ref{q_vis_XC97}), respectively (see Appendix \ref{enegy_xu_chen_appendix} for
more details). Based upon the global picture of the accretion flow, inflow happens around 
the equatorial plane while the outflow does appear at high latitudes. 
To show that the range of the density index, only for highly non-relativistic cases, must be $ n < 3/2 $, in what follows, 
we investigate the energy equation in inflow and wind regions in details.

\subsection{Energy equation in the inflow region} \label{energy_inflow_region}

As proved by \citealt{Mihalas and Mihalas 1984}, the viscous dissipation term 
is always positive everywhere, $  Q_\mathrm{vis} > 0 $. 
Besides, the symmetry boundary conditions dictate that the latitudinal component 
of the velocity should be null at the equatorial plane, i.e., $ v_{\theta} (\pi/2) = 0 $ 
(see equation (\ref{boundary_conditions})) which means that the first term of the equation 
(\ref{q_adv_XC97}) is dominated at midplane. Necessarily, this term must be positive 
to satisfy the energy equation,

\begin{equation} \label{term1_q_vis_inflow}
	 \left[ n  - \frac{1}{ \gamma - 1} \right] p_\mathrm{g} v_{r}  > 0,
\end{equation}

where $ \gamma $ is the adiabatic index. Since the gas pressure is always positive and $ v_{r} < 0 $ at the equatorial plane, then 
the above equation will be satisfied there only if,

\begin{equation} \label{coef_q_vis}
n  - \frac{1}{ \gamma - 1} < 0.
\end{equation}
 
As mentioned in the introduction, numerical simulations of the hot accretion flow 
suggest that in the non-relativistic cases, $ \gamma $ is very close to $ 5/3 $. 
Consequently, to satisfy the above equation, with $ \gamma = 5/3 $, the density 
index must be\footnote{It should be emphasized here that this analysis is very specific since we adopt self-similar approximation
with constant Mach number along the radial direction and $\gamma=5/3$ which is only applicable for highly non-relativistic cases.},
\begin{equation} \label{density_index_n}
 	n < \frac{3}{2}. 
\end{equation}
This value for n clearly proves that the bipolar ejection outflow solution of XC97 is not physical where they set n = 2.5 
(see section 3 and also panel (b) of figure 1 in XC97).
\subsection{Energy equation in the wind region}

At high latitudes, if wind exists and launches, the radial velocity must 
be positive ($ v_{r} > 0 $). At the rotation axis, we have similar boundary condition for the latitudinal 
component of the velocity, i.e., $ v_{\theta}(0) = 0 $. Additionally, as proved from the energy 
equation in the inflow region, the density index must be $ n < 3/2 $. Therefore, the advection
term at the pole must be negative as,

\begin{equation} \label{term1_q_vis_outflow}
	 \left[ n  - \frac{1}{ \gamma - 1} \right] p_\mathrm{g} v_{r}  < 0.
\end{equation}
Since the viscous heating term of the energy equation is always positive, 
to satisfy the above equation, we need an additional term in the 
right-hand side of the energy equation which allows for the positive 
radial velocity near the pole. 
TM06 and KS13 showed that thermal conduction can be negative enough at high 
latitudes to overcome positive sum of the viscous heating terms and generate positive nonzero 
radial velocity about the rotation axis.

In essence, in hot accretion flows with very low mass accretion rates, 
the electron mean-free path is much larger than electron gyro-radius.
Hence, Coulomb collisions are not exceptional. 
For instance, in our Galactic Center, Sgr $ \mathrm{A^*} $, with mass accretion 
rate $ \dot{\mathrm{M}} \approx 10^{-7} - 10^{-8}\, \mathrm{M}_{\odot} \mathrm{yr^{-1}} $, 
the electron mean-free path
$ l \sim 1.3 \times 10^{17} \mathrm{cm} $ and the gyro-radius of electrons
$ R_\mathrm{gyro} (= m_\mathrm{e} v_\mathrm{th} c / q B) \sim 10^5 \mathrm{cm} $ 
($ m_\mathrm{e} $, $ v_\mathrm{th} $, $ c $, $ q $ and $ B $, are electron mass, 
thermal speed of electron, speed of light, electron charge, and magnetic field, respectively).
Thermal conduction can play a striking role in such a system as 
can make a heat flux from hot inner regions to the cold outermost of the flow. 
The gas in outer region has the possibility to be heated to a temperature 
higher than the local virial temperature. This increase in the temperature 
can drive thermal outflow/wind and consequently decrease the mass accretion rate.

In this paper, we will solve the HD equations of the hot accretion flow with thermal conduction in the whole $ \theta $ direction. 
We will also consider all components of the velocity and the viscous stress tensor and impose the boundary conditions at both the rotation axis and the equatorial plane.

\section{Basic Equations and Assumptions} \label{sec:basic_equations}

The HD equations of the hot accretion flow with thermal conduction 
can be written as,

\begin{equation} \label{eq:continuity}
  \frac{ \mathrm{d} \rho}{\mathrm{d} t} +\rho \nabla \cdot \bm{v} = 0,
\end{equation}
\begin{equation}\label{eq:momentum}
\rho  \frac{\mathrm{d} \bm{v}}{\mathrm{d} t} = - \rho \nabla \psi  - \nabla p + \nabla \cdot \bm{\sigma},
\end{equation}

\begin{equation}\label{eq:energy}
	Q_\mathrm{adv}  =  Q_\mathrm{vis} + Q_\mathrm{c},
\end{equation}
where, $ \rho $ is the mass density, $ \bm{v} $ is the velocity, 
$ \psi \left[ = - GM / r \right] $ is the Newtonian potential (where 
$ r $ is the distance from central black hole, $ M $ is the black 
hole mass, and $ G $ is the gravitational constant),
$ p $ is the gas pressure, $ \bm{\sigma} $ is the viscous stress tensor, and the
$ \mathrm{d}/\mathrm{d} t \equiv \partial / \partial t +  \bm{v} \cdot \nabla $ denotes
the Lagrangian or comoving derivative. In the equation (\ref{eq:energy}),  $ Q_\mathrm{adv} $,
$ Q_\mathrm{vis}  $, and $ Q_\mathrm{c} $ are advection, viscous and thermal conduction
terms of the energy equation, respectively, which can be described as,

\begin{equation} \label{Q_adv}
	Q_\mathrm{adv} = \rho \frac{\mathrm{d} e}{\mathrm{d} t} - \frac{p}{\rho} \frac{\mathrm{d} \rho}{\mathrm{d} t}, 
\end{equation}

\begin{equation} \label{Q_vis}
	Q_\mathrm{vis} =  f \nabla \bm{v} : \bm{\sigma},
\end{equation}

\begin{equation} \label{Q_c}
	Q_\mathrm{c} = - \nabla \cdot \bm{F}_\mathrm{c}. 
\end{equation}

Here, $ e $ is the internal energy of the gas, and $ \bm{F}_\mathrm{c} $ is 
the heat flux due to the thermal conduction. We adopt the adiabatic equation of state
as $ p =  \left( \gamma - 1 \right) \rho e $. In purely HD limit, $ \bm{F}_\mathrm{c} $ is defined as,
\begin{equation} \label{F_conduction}
	\bm{F}_\mathrm{c} = - \chi \nabla T,
\end{equation}
where $ \chi $ is the thermal diffusivity and $ T $ is the gas temperature. 
In one-temperature structure, as our case, $ T $ can be written as,
\begin{equation} \label{temperature}
	T = \frac{\mu m_{p}}{k_{\!_{B}}} \frac{p}{\rho},
\end{equation}
where, $ \mu $, $ m_{p} $, and $ k_{\!_{B}} $ are
mean molecular weight, proton mass, and Boltzmann constant, respectively. 
Numerical simulations of \citealt{Sharma et al. 2008} and \citealt{Bu et al. 2016}), 
assumed $ \kappa = \chi T/p \equiv \alpha_\mathrm{c} \left(GM r\right)^{1/2} $
with the dimensionless conductivity takes the value of $ \alpha_\mathrm{c} \simeq (0.2-2) $.
Here, we follow those numerical simulations by adopting the same definition for thermal conduction and consider the same range for $ \alpha_{c} $ except in section \ref{Dependency_of _the_Solutions} we consider small value of $ \alpha_c $ to check the dependency of the results to this parameter. The viscous stress tensor is given by,

\begin{equation}\label{viscous_stress_tensor}
	\sigma_{ij} =  \rho \nu \left[ \left( \frac{\partial v_{j}}{\partial x_{i}} 
	+ \frac{\partial v_{i}}{\partial x_{j}} \right)  
	- \frac{2}{3} \left( \nabla \cdot \bm{v} \right) \delta_{ij} \right],
\end{equation}
where $ \nu $ is called the kinematic viscosity coefficient and $ \delta_{ij} $ is 
the usual Kronecker delta. Note that the bulk viscosity is neglected here. 
It is well known that in real accretion flow, the angular momentum is transferred 
by Maxwell stress associated with MHD turbulence 
driven by MRI (\citealt{Balbus and Hawley 1998}). 
In our current HD case, we approximate the effect of the magnetic stress by 
adding viscous terms in momentum and energy equations (see e.g., \citealt{Yuan et al. 2012b}). 
The kinematic viscosity coefficient is calculated with the $ \alpha $-prescription 
(\citealt{Shakura and Sunyaev 1973}) as,

\begin{equation}\label{kinematic_viscosity_coefficient}
	\nu =  \alpha \frac{p}{\rho  \Omega_{\!{_K}}},
\end{equation}
where $ \Omega_{\!{_K}} \equiv (GM/r^{3})^{1/2}$ is the Keplerian angular velocity and $ \alpha $ is
the viscosity parameter. We use spherical coordinates $ (r, \theta, \phi) $ to solve the full set 
of equations including all viscous terms. The disc is taken to be axisymmetric and 
steady state. By implementing all the above mentioned assumptions and definitions into equations
(\ref{eq:continuity})-(\ref{eq:energy}), we obtain the partial differential equations (PDEs)
presented in Appendix \ref{pde_appendix} (see equations (\ref{pde_continuity})-(\ref{pde_energy}) 
for more details). 

The global numerical simulations of hot accretion flows show that the physical variables 
of the flow can be described by power-law function of radius far away from the radial boundaries.
For instance, the radial profile of the density follows $ \rho(r) \propto r^{-n} $ with 
$ n < 3/2 $ ( see, e.g., \citealt{Stone et al. 1999}; \citealt{Yuan et al. 2012a, Yuan et al. 2012b, Yuan et al. 2015}).
In order to solve equations (\ref{pde_continuity})-(\ref{pde_energy}) by numerical methods, 
we impose self-similar solutions to remove the radial dependency of the variables.
To do so, we introduce the fiducial radius $ r_{\!_{0}} $ with the self-similar solutions 
as a power-law form of $ (r/r_{\!_{0}}) $. Accordingly, the physical variables of the hot accretion flow 
will be written in the following forms,

\begin{figure*}[ht!]
\includegraphics[width=\textwidth]{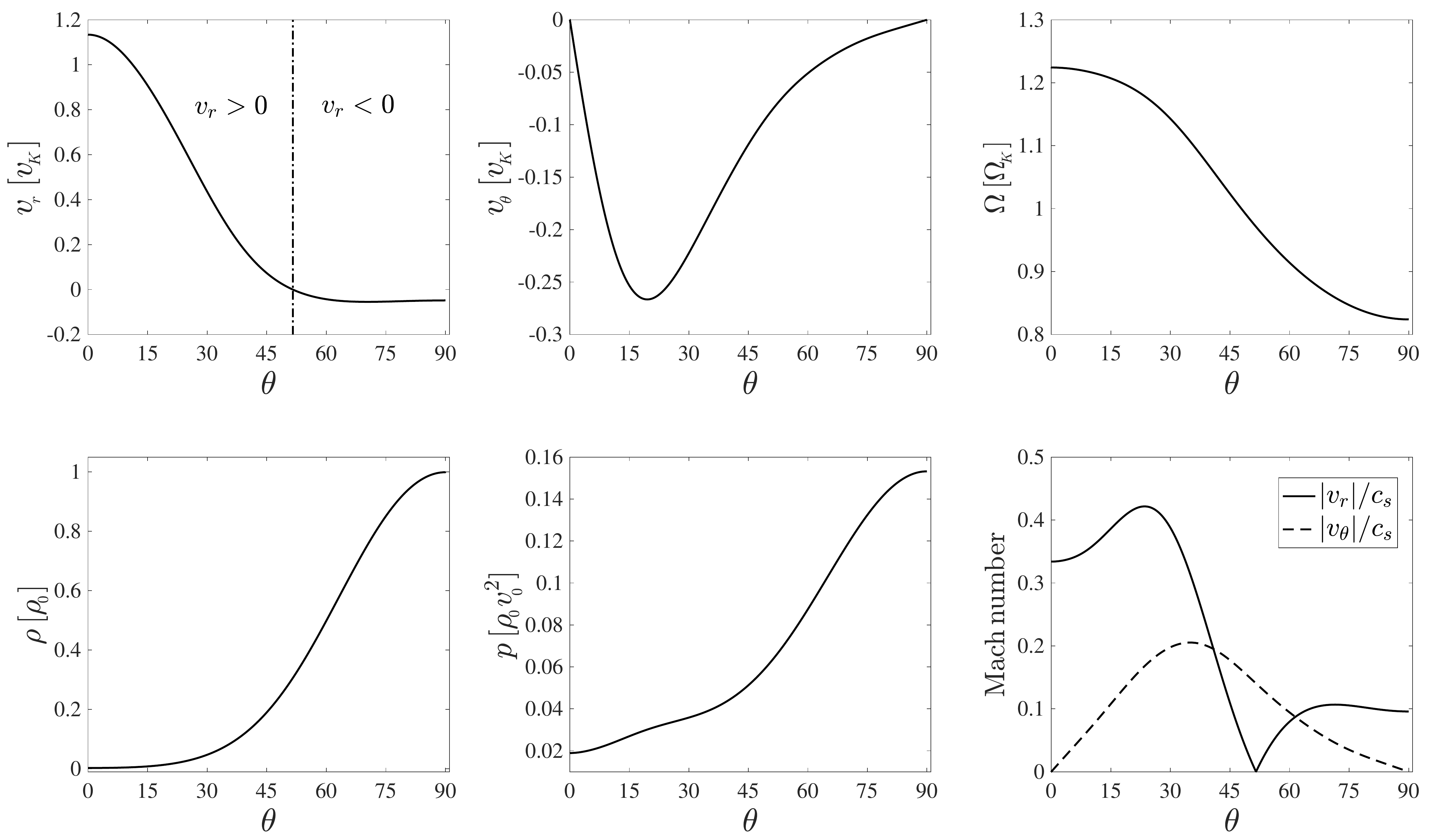} 
\caption{Latitudinal profile of the physical variables of the fiducial model. 
\textit{Top row:} radial velocity in the unit of Keplerian velocity, $ v_{\!_{K}} $;
the dash-dotted line shows the location of $ v_{r} = 0 $ that is about $ 52^{\circ} $ (left panel); 
latitudinal velocity, $ v_{\theta} $ (middle panel); angular velocity in the unit 
of Keplerian angular velocity, $ \Omega_{\!_{K}} $ (right panel).
\textit{Bottom row:}  
density in the unit of density of mid-plane at $ r_{\!_{0}} $, i.e., $ \rho_{\!_{0}} $ (left panel); 
gas pressure in the unit of $ \rho_{\!_{0}} v_{\!_{0}}^{2} $ where $ v_{\!_{0}} (\equiv \sqrt{GM/r_{\!_{0}}}) $ 
is the Keplerian velocity at $ r_0 $ (middle panel); Mach numbers (right panel).
\label{physical_variables}}
\end{figure*}

\begin{equation}\label{rho_selfsimilar}
  \rho \left(r, \theta \right) = \rho_{\!_{0}} \left( \frac{r}{r_{\!_{0}}} \right)^{-n} \bar{\rho}(\theta) ,
\end{equation}

\begin{equation}\label{vr_selfsimilar}
  v_{r} (r, \theta) = v_{\!_{0}} \left( \frac{r}{r_{\!_{0}}} \right)^{-1/2} \bar{v}_{r}(\theta), 
\end{equation}

\begin{equation}\label{vtheta_selfsimilar}
  v_{\theta} (r, \theta) = v_{\!_{0}} \left( \frac{r}{r_{\!_{0}}} \right)^{-1/2} \bar{v}_{\theta}(\theta),
\end{equation}

\begin{equation}\label{vphi_selfsimilar}
  v_{\phi} (r, \theta) = v_{\!_{0}} \left( \frac{r}{r_{\!_{0}}} \right)^{-1/2} \bar{\Omega}(\theta) \sin(\theta),
\end{equation}

\begin{equation}\label{p_gas_selfsimilar}
  p (r, \theta) = p_{\!_{0}} \left(  \frac{r}{r_{\!_{0}}} \right)^{-n-1} \bar{p}_\mathrm{g} (\theta),
\end{equation}
where $ r_{\!_{0}} $, $ \rho_{\!_{0}} $, $ v_{\!_{0}} \left[= \sqrt{GM/r_{\!_{0}}} \right]$, and 
$ p_{\!_{0}} \left[= \rho_{\!_{0}} v_{\!_{0}}^2 \right] $, are the units of length, density, velocity, 
and gas pressure, respectively. Substituting the above 
self-similar solutions into equations (\ref{pde_continuity})-(\ref{pde_energy}), the
radial dependency will be removed and the system of PDEs will 
be reduced to a set of ordinary differential equations (ODEs) presented in 
Appendix \ref{ode_appendix}. 
The ODE equations (\ref{ode_continuity})-(\ref{ode_energy}) consist of five physical variables:
$ v_{r} (\theta) $, $ v_{\theta} (\theta) $, $ \Omega (\theta) $, 
$ \rho(\theta) $, $ p_\mathrm{g}(\theta) $ and also their first and second 
derivatives. As we mentioned in introduction, the integration will not stop at some angle near the rotation axis
(see e.g., \citealt{Jiao and Wu 2011}, \citealt{Mosallanezhad et al. 2016}, \citealt{Samadi and Abbassi 2016}). Instead the computational domain will be extended 
from the rotation axis, $ \theta = 0 $, to the equatorial plane, $ \theta = \pi/2 $.
Following previous analytical solutions of hot accretion flow, e.g., NY95 
and TM06, all physical variables are assumed to be even symmetric, continuous, 
and differentiable at both boundaries. Since we also include the latitudinal component of the velocity, 
$ v_{\theta} $, its value will be null at both the equatorial plane and the rotation axis. 
Thus, the following boundary conditions at $ \theta = 0 $ and $ \theta = \pi/2 $ will be imposed: 

\begin{equation} \label{boundary_conditions}
	\frac{\mathrm{d} \rho}{ \mathrm{d} \theta} = \frac{\mathrm{d} p_\mathrm{g}}{\mathrm{d} \theta} 
	=  \frac{\mathrm{d} \Omega}{\mathrm{d} \theta} = \frac{\mathrm{d} v_{r}}{ \mathrm{d} \theta} =  v_{\theta} = 0.
\end{equation}

To satisfy all boundary conditions at both ends, the relaxation method will be adopted 
mainly because the set of ODEs have extraneous solutions, and also there exists 
singularity at the rotation axis. 
To have a good resolution at both sides where the boundary conditions set, 
we divide the $ \theta $ direction into 5000 grids with stretch grid as follow: 
From $ \theta = 0 $ to $ \theta = \pi/4 $ the grid size ratio is set as 
$ \mathrm{d}\theta_\mathrm{i+1}/ \mathrm{d}\theta_\mathrm{i} = 1.003 $
while from $ \theta = \pi/4 $ to $ \theta = \pi/2 $ the grid size ratio is set as 
$ \mathrm{d}\theta_\mathrm{i+1}/ \mathrm{d}\theta_\mathrm{i} = 0.997 $.
We calculate the variables at the cell center of these grids.
The absolute error tolerance is set to $ 10^{-15} $. The most difficult part 
of solving the set of equations is providing an appropriate guess for the required solutions.
In this study, we use \textit{Fourier cosine series} for the initial guess of all physical variables except
for $ v_{\theta}(\theta) $ which we used \textit{Fourier sine series}.
Since our solutions are satisfying the boundary conditions at both the rotation axis as well as 
the equatorial plane, it does assure us that well-behaved solutions will be derived in the whole $ \theta $
direction. In the next section we will explain the behaviors of the physical variables in detail.

\section{Numerical Results} \label{sec:results}

\subsection{The solutions of the fiducial model} \label{sec:solutions}

\begin{figure*}[ht!]
\includegraphics[width=0.5\textwidth]{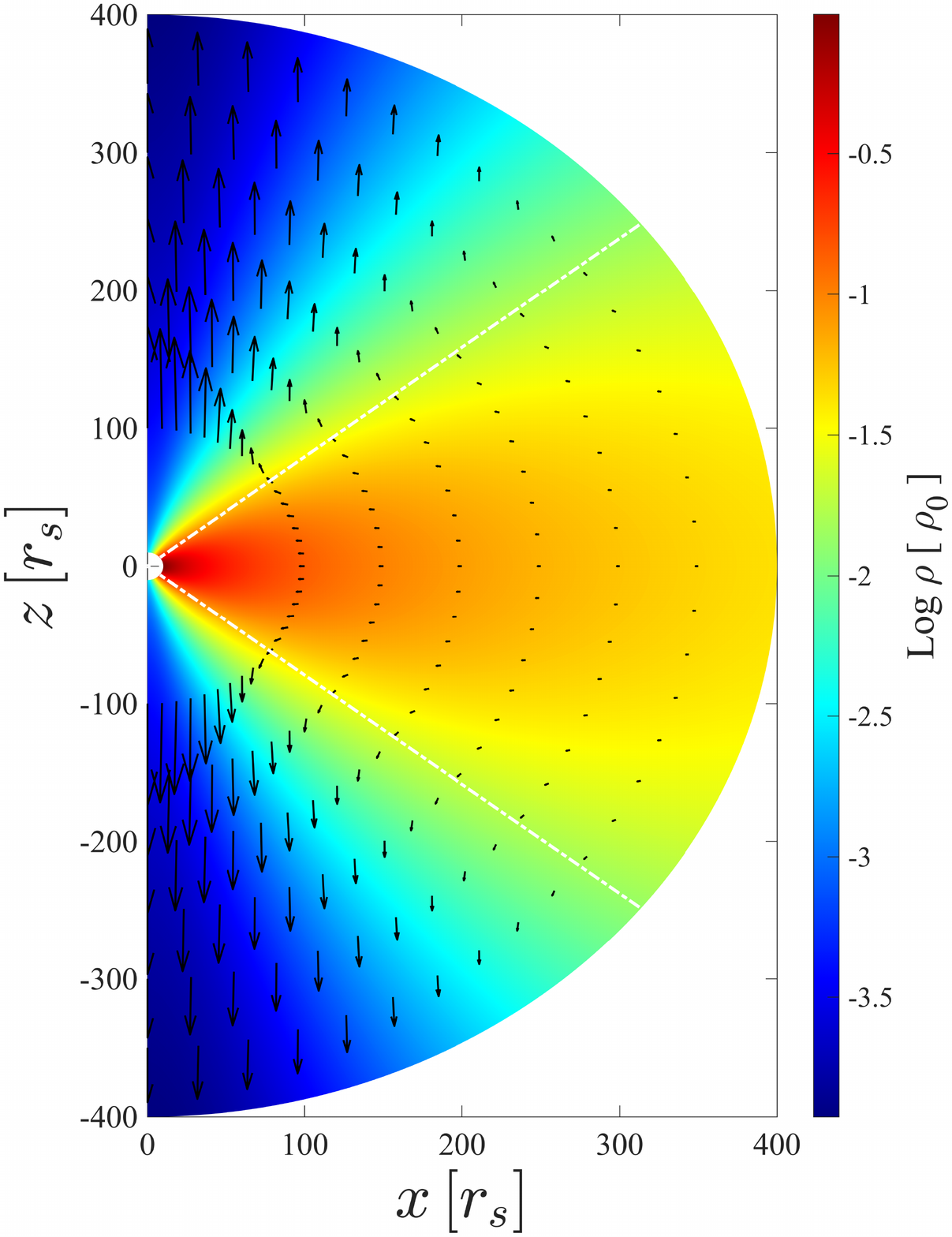} 
\includegraphics[width=0.5\textwidth]{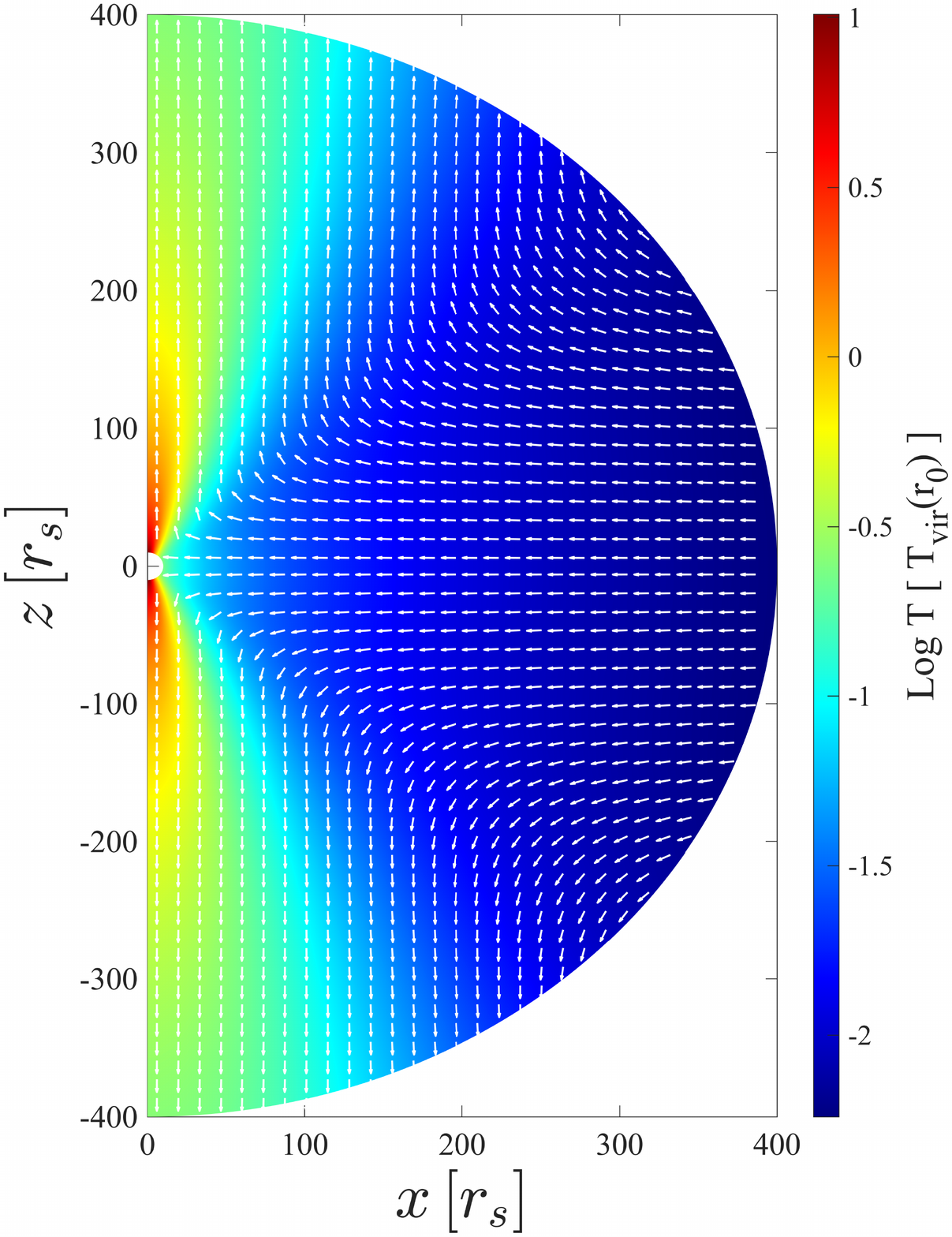}
\caption{Two-dimensional distribution of the density (left panel) and the temperature (right panel) 
based on the self-similar solutions. Both panels are overlaid with the poloidal velocity, 
$ \vec{v}_{p} = v_{r} \hat{\bm{r}} + v_{\theta} \hat{\bm{\theta}} $ . In the left panel, 
the arrows are plotted only at six different radii in the unit of schwarzschild radius, i.e., 
$ r = [100, 150, 200, 250, 300, 350]\, r_\mathrm{s} $ to show the strength of the outflow. 
The white dash-dotted lines show the location of $ v_{r} = 0 $. In the right panel, 
the poloidal velocity is normalized with its absolute value, $ | \vec{v}_{p} | = \sqrt{ v_{r}^{2} + v_{\theta}^2 } $
to denote the direction of the vectors. Here, $ T_\mathrm{vir}(r_{\!_{0}}) = GM m_{p} / ( 3 k_{B} r_{\!_{0}} ) $
is the virial temperature at $ r_{\!_{0}} = 10\, r_\mathrm{s} $.
 \label{density_temperature}}
\end{figure*}

To solve the system of ODEs numerically, we integrate the equations 
(\ref{ode_continuity})-(\ref{ode_energy}) from the rotation axis
to the equatorial plane to get the latitudinal profiles of all 
physical variables of our fiducial model shown in Figure \ref{physical_variables}. 
The parameters are set as $ \alpha = 0.15 $, $ \alpha_c = 0.2 $, $ n = 0.85 $, $ f = 1 $, and 
$ \gamma = 5/3 $. The velocities are scaled with the Keplerian velocity at $ r = r_{\!_{0}} $, 
i.e., $ v_{\!_{0}} = \sqrt{GM/r_{\!_0}} $. The density is normalized with $ \rho_{\!_{0}} $, 
and the pressure is also normalized with $ \rho_{\!_{0}} v_{\!_{0}}^{2} $ at $ r_{\!_{0}} $. 
We assumed the radius $  r_{\!_{0}} $ is located at the equatorial plane where 
the maximum density of the accretion flow
is accumulated there. In top left panel of Figure \ref{physical_variables}, 
$ v_{r} = 0 $ shows the inclination of the inflow to the wind region which is around 
$ \theta = 52^{\circ} $\footnote{We made a parameter study and found that inclination of the inflow to the wind region does not change too much with different set of input parameters. The changes are only in the range of $ \theta \sim 50^{\circ} - 55^{\circ}$.}. As it is seen, $ v_r $ is negative in the inflow region where the 
matter goes toward the BH while in the wind region it becomes positive and 
reaches a value slightly above Keplerian velocity of that radius. 
In the top middle panel, $ v_{\theta} $ is negative in the vertical direction from 
the equatorial plane toward the rotation axis. Furthermore, due to the boundary conditions 
it is zero in both boundaries and minimum around $ \theta \simeq 20^\circ $.
In top right panel, the angular velocity tendency, as it increases from the 
inflow region toward the wind region, shows that the angular momentum is transported 
away from the system by the wind. Moreover, the angular velocity exceeds Keplerian velocity
around $ \theta \simeq 45^\circ $ and is quite super-Keplerian in the wind region.
In bottom left panel, the concentration of the density is at the equatorial plane and then drops toward 
the wind region so that reaches its minimum value around the rotation axis. The trend 
of the gas pressure, bottom middle panel, is also the same for the density, 
maximum at the equator and the minimum at the rotation axis. 
From the bottom right panel, $ v_r $ and $ v_{\theta} $ both remain 
subsonic in whole $ r-\theta $ domain. The main reason is that the density drops faster 
than the pressure from equatorial plane toward the rotation axis. Therefore, the sound speed,
$ c_\mathrm{s} = \sqrt{\gamma p / \rho} $, or equivalently the gas temperature increases as
$ \theta $ angle decreases. This panel clearly shows that both $ | v_{r} | / c_\mathrm{s}  $ 
and $ | v_{\theta} | / c_\mathrm{s} $ do not pass through a sonic point (where the singularity exists)
at small angles. The behavior of Mach number for $ v_{r} $ is that 
it first drops and then increases going from $ \theta = 52^{\circ} $ (where the radial velocity is zero, $ v_{r} = 0 $) 
toward the rotation axis. 
Also, $ | v_{\theta} | / c_\mathrm{s} $ first increases until around $ \theta \simeq 35^{\circ} $ 
and then decreases toward the rotation axis, and this is because we plot the absolute value of 
Mach numbers. The overall behavior of the physical variables are almost the same as those in 
XC97, TM06, and KS13. More precisely, the similarity can be seen in the inflow-wind structure, 
inflow with negative radial velocity around the equatorial plane and wind with positive radial velocity
around the rotation axis. It is very diagnostic in the density profile which
drops from the equatorial plane to the rotation axis in all above studies showing a disc-shape structure 
(see also left panel of Figure \ref{density_temperature}). Since in the present study we include $ v_{\theta} $, 
we can easily compare our results with XC97 and KS13. For instance, the wind radial velocity is much more higher
than the inflow radial velocity at high latitudes which is exactly similar to the Figure 2 of XC97 and also Figure 1 of KS13\footnote{Note here that the main differences between our solution and the accretion outflow solution of XC97 are; (1) we include thermal conduction, and (2) the value of the radial velocity at 
the equatorial plane is self-consistently determined here,
while in XC97, it is fixed as $  v_{r}(\pi/2) = -0.05 v_{\!_{K}} $.}. 
The trend of the angular velocity in the present work increasing 
toward the rotation axis is similar to that in both XC97 and KS13, 
which helps the wind to accelerate with high speed and cause the angular momentum 
is transferred outward (see Figure 2 of XC97 and also the bottom panel of figure 2 in KS13). 
However, in figure 2 of TM06, angular velocity, $ \Omega $, decreases with decreasing $ \theta $. 
One of the differences of the present work and XC97 with KS13 is that the angular velocity becomes 
super-Keplerian at small angles. Another one is U-shape behavior of $ v_{\theta} $,
it first falls down and then goes up toward the rotation axis, 
while it keeps increasing continuously toward the opening angle in KS13. 
The reason can be interpreted due to different assumptions which these studies have made 
in their calculations. In addition, we and XC97 both put the second boundary condition at the rotation axis where 
the gradient of the physical variables as well as $ v_{\theta} $ must be zero. 
Nevertheless, KS13 started the integration from the equator and went to a certain inclination 
where the physical constraint was satisfied and considered this inclination as the upper boundary of the accretion flow.
Overall, due to appropriately imposing physical boundary conditions at the rotation axis as well as considering thermal conduction in this study, our results show strong wind at high latitudes.

\begin{figure*}[ht!]
\includegraphics[width=\textwidth]{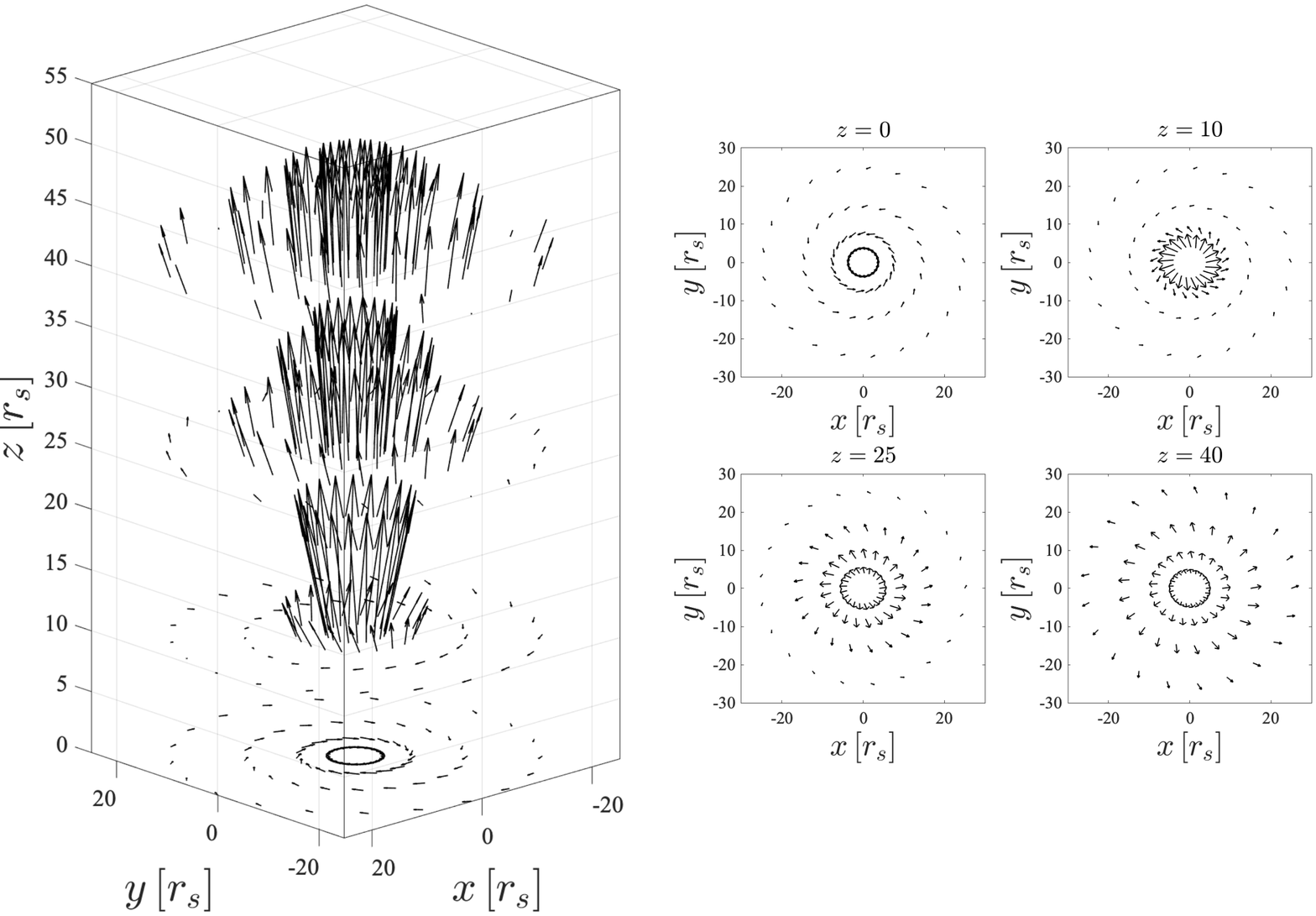} 
\caption{Velocity field in three dimensional space (left plot), in four cylindrical radii, i.e., 
$ R = [4, 8, 15, 25]\, r_\mathrm{s} $ and also in four different heights, i.e., $ z =  [0, 10, 25, 40]\, r_\mathrm{s} $. 
The two dimensional velocity fields have been plotted at four different heights (four right panels).  
\label{velocity_field}}
\end{figure*}

\begin{figure*}[ht!]
\includegraphics[width=0.9\textwidth]{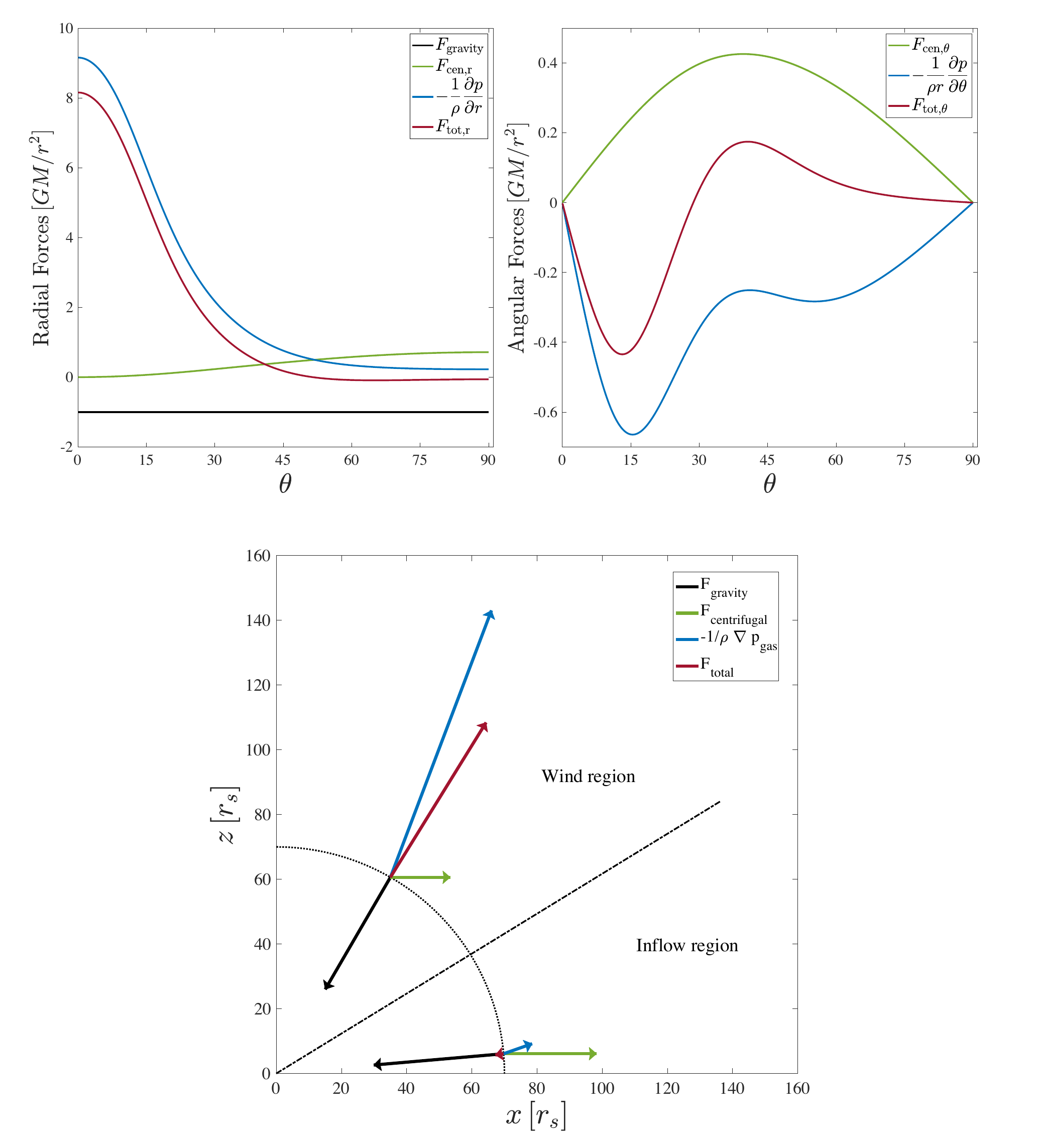} 
\centering
\caption{Angular distributions of the radial forces (top left panel) and angular forces (top right panel) in the unit of gravitational force, and the force analysis in the inflow/wind region to show the driving mechanism of the wind at $ r = 70 \, r_{s} $ (bottom panel). 
The forces include gravity (black), centrifugal force (green), gradient of gas pressure (blue), and their sum (red). 
In the bottom panel, the length of the arrows schematically denote the magnitude of the forces while the direction of arrows show that of the forces. The dash–dotted line shows the location of $ v_{r} = 0 $, and the dotted line is for the radius where the forces are calculated. 
The forces are calculated at two representative locations in wind and inflow regions, $ \theta = 30^{\circ} $ and $ \theta = 85^{\circ} $, respectively.
\label{force_analysis}}
\end{figure*}

The two dimensional inflow-wind structure of the flow is shown in Figure \ref{density_temperature}, 
in the density and temperature contours. The left panel is overlaid with the poloidal velocity
at six different radii in the unit of Schwarzchild radius, i.e., $ r = [100, 150, 200, 250, 300, 350]\, r_\mathrm{s} $
to show the strength of the wind.  In the right panel, the poloidal velocity is normalized with 
its absolute value, $ | \vec{v}_{p} | = \sqrt{ v_{r}^{2} + v_{\theta}^2 } $, which denotes the 
direction of the vectors. As it can be seen from velocity field, around mid-plane the flow is accreted 
toward the BH until the white dash-dotted line ($ v_r = 0 $) in the left panel. 
From this line to the high latitudes the velocity vector deflects outward and the flow escapes away.
Moreover, it is clear that the higher the latitudes, the stronger the wind is. It is also noted that 
the density and temperature profiles are plotted based on 
our self-similar solutions with density index $ n = 0.85 $ (see equations (\ref{rho_selfsimilar})-(\ref{p_gas_selfsimilar})). 
From this figure, the maximum amount of the density is located at the inner region of the equatorial plane. So, as 
we go to outer regions and small $ \theta $ angles, the density decreases rapidly and reaches its minimum in the disc.
The density contour is identical to that plotted in the top left panel of figure 2 in XC97 for 
the accretion outflow solution. The torus-like shape of the density is also in agreement with the numerical simulations of hot accretion flow (see e.g., \citealt{Yuan et al. 2012b}). 
From the right panel, the minimum temperature belongs to the inflow region inside the disc, 
while the wind region has the highest temperature which is much less dense than the disc. 
Indeed, at high latitudes, thermal conduction heats up the flow then, the temperature goes up and leads to a rapid change of the gradient of the gas pressure, launching the thermal wind (see also top left panel of Figure \ref{force_analysis}).

To show the strength of the wind, we also plot three dimensional (3D) 
velocity field in Figure \ref{velocity_field} in four cylindrical radii, i.e., 
$ \mathrm{R} = [4, 8, 15, 25]\, r_\mathrm{s} $, 
and also in four different heights, $  z = [0, 10, 25, 40]\, r_\mathrm{s} $ (left panel). 
From the 3D figure, it is clear that the flow is purely inflow at the equatorial 
plane of the disc, $ z = 0 $. In higher $ z $, the velocity vectors
are not parallel to the equatorial plane and deflected away.  
In addition, arrows become stronger around the rotation axis at small heights as well as small radii. 
For the better view of the velocity field, 
the two dimensional (2D) plots in four different heights in $ x-y $ plane 
are shown in the right panel of Figure \ref{velocity_field}.

It is of interest to know which mechanism is dominant to drive wind in the 
hot accretion flow in HD case. For this purpose, in Figure \ref{force_analysis}, we 
analyze all forces in the unit of the gravitational force, including the gradient 
of gas pressure (blue), the centrifugal force (green), the gravitational force (black), 
and the sum of the forces (red). It is also worthwhile to find the prominent radial 
and angular components of the forces in the inflow and wind regions. 
In the top left panel, the radial components of the forces are shown. 
Since the gravitational force only changes with radius, in this panel, it has 
a fixed negative value, i.e., $ F_\mathrm{gravity} = -1 $, in whole $ \theta $ direction.
In the inflow region, the gravity is prevailing while from $ \theta \simeq 52^\circ $ upward, 
the gradient of the gas pressure becomes dominant (wind region).
As it can be seen, the sum of the radial components of the gradient of the gas pressure 
and the centrifugal force cannot dominate the gravity in the inflow region 
so, the total force remains negative there. 
This is also clear from the bottom panel of Figure \ref{force_analysis}, 
where the forces are plotted at the region near the equatorial plane, i.e., $ \theta = 85^{\circ} $,
since the latitudinal components of the forces are almost negligible.
The top right panel of this figure shows that the vertical component of the centrifugal force 
is always positive and dominant in the inflow region. On the other hand, in the wind region, 
the $ \theta $ component of the gradient of the gas pressure is dominated and the sum of these forces
in the vertical direction becomes negative. 
In bottom panel of Figure \ref{force_analysis}, the forces are calculated at two representative locations 
in wind and inflow regions, $ \theta = 30^{\circ} $ and $ \theta = 85^{\circ} $, respectively.
The dash-dotted line represents the barrier between inflow and wind regions
corresponded to $ v_r = 0 $ and the dotted line is for the radius where the forces 
are evaluated, i.e., $ r = 70\, r_\mathrm{s} $.
This panel clearly shows that in the inflow region gravity is dominant so the matter moves 
inward, and in the wind region the sum of the forces is outward due to the strong value of the 
gradient of the gas pressure. These results are in agreement with those presented in
numerical HD simulations of the hot accretion flow (see \citealt{Yuan et al. 2012b}). 

\begin{figure}[ht!]
\includegraphics[width=0.48\textwidth]{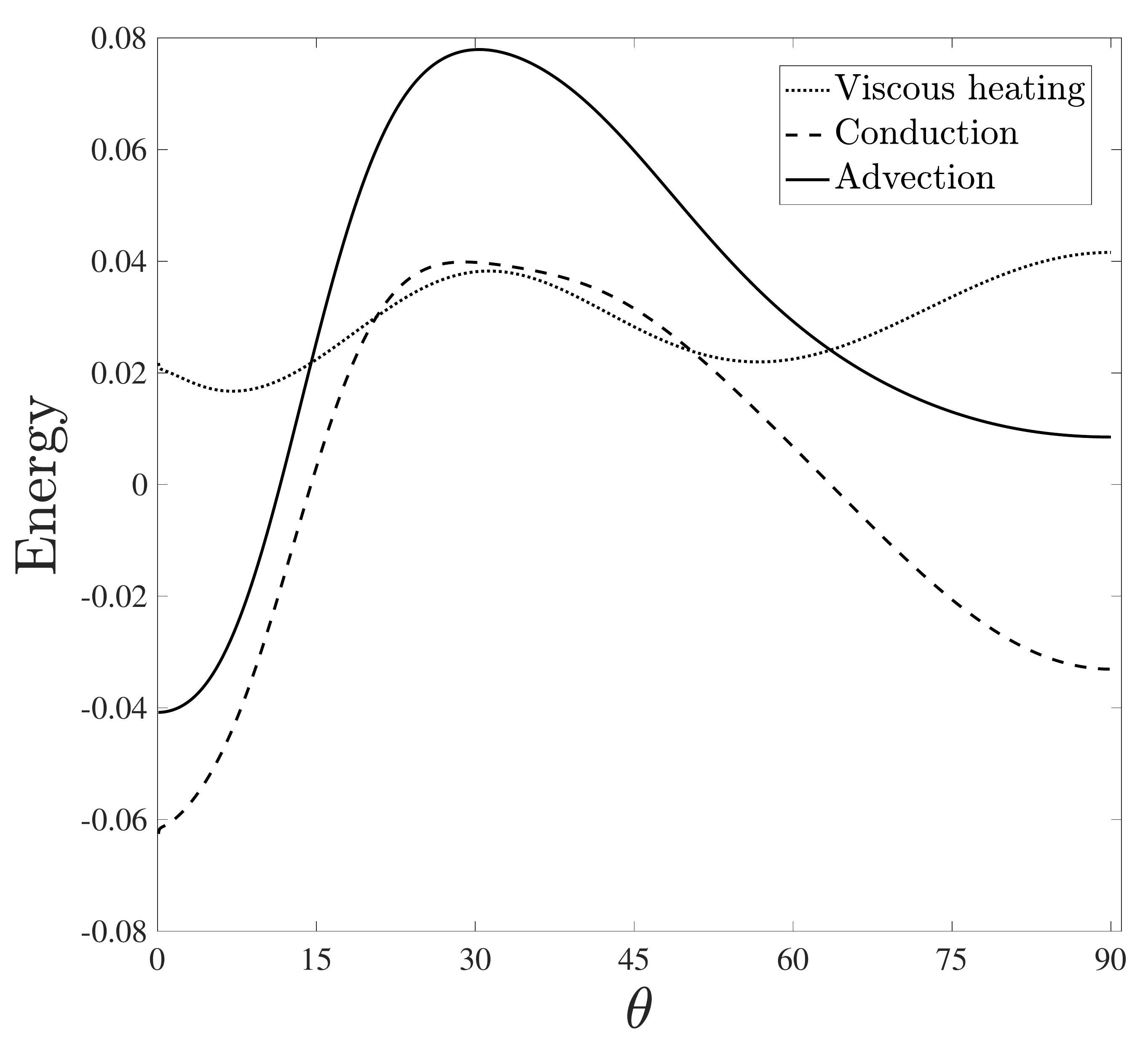}
\centering
\caption{Latitudinal profiles of the viscous heating (dotted line), the conduction (dashed line), 
and the advection (solid line) terms of energy equation presented in equation (\ref{ode_energy}).
\label{energy}}
\end{figure}

Figure \ref{energy} shows the latitudinal profile of three terms of energy equation including 
viscous heating (dotted line), conduction (dashed line), and advection (solid line) based on the solutions 
of the fiducial model. This figure shows that in the inflow region the viscosity and advection terms 
have positive values while the thermal conduction has a negative value,
so the advection can cool the flow in the disc.
At the polar region, the thermal conduction is much more negative 
and the viscous heating term is almost negligible. Consequently, the sum of these two terms permits 
a negative advection term near the rotation axis which means the advection acts to heat 
the flow and produce wind. Note that the latitudinal profile of three terms 
of the energy equation are not identical to the ones presented in TM06 because of 
two main reasons; (1) the thermal conduction term defined here is not similar to that written in TM06
(we follow the numerical simulation of \citealt{Sharma et al. 2008} since the TM06 definition of the thermal conduction 
is not consistent with the self-similarity adopted here\footnote{In TM06, since the density was proportional to $ \propto r^{-3/2} $,
the thermal conductivity coefficient
defined as $ \lambda(r) = \lambda_{0} r^{-1} $ to preserve the radial self-similarity of the solutions.
Here, to satisfy the radial self-similar solutions, 
we cannot follow TM06 definition for thermal conduction since the density is proportional to $ \propto r^{-n} $.}). 
(2) The latitudinal component of the velocity, $ v_{\theta} $, 
exists in both advection and viscous terms of our equations, while this component of the velocity was ignored in TM06.

\subsection{Bernoulli parameter} \label{Bernoulli:!}

In almost all previous analytical solutions of the hot accretion flow, the Bernoulli parameter 
was calculated to show the existence of the wind/outflow.
The Bernoulli parameter is defined as the sum of the kinetic energy,
the potential energy and the enthalpy of the accreting gas. 
In fact, the Bernoulli parameter has been of significant concern as it shows 
whether wind or outflow would probably emanate from the accretion flow (\citealt{Narayan and Yi 1995}).
The Bernoulli parameter can be defined as,

\begin{equation} \label{Bernoulli_parameter}
	Be = \frac{1}{2} \bm{v}^{2} + h + \psi,
\end{equation}
where, $ h = \gamma p / \left[ \rho (\gamma - 1) \right] $ is the enthalpy.

\begin{figure}[ht!]
\includegraphics[width=0.49\textwidth]{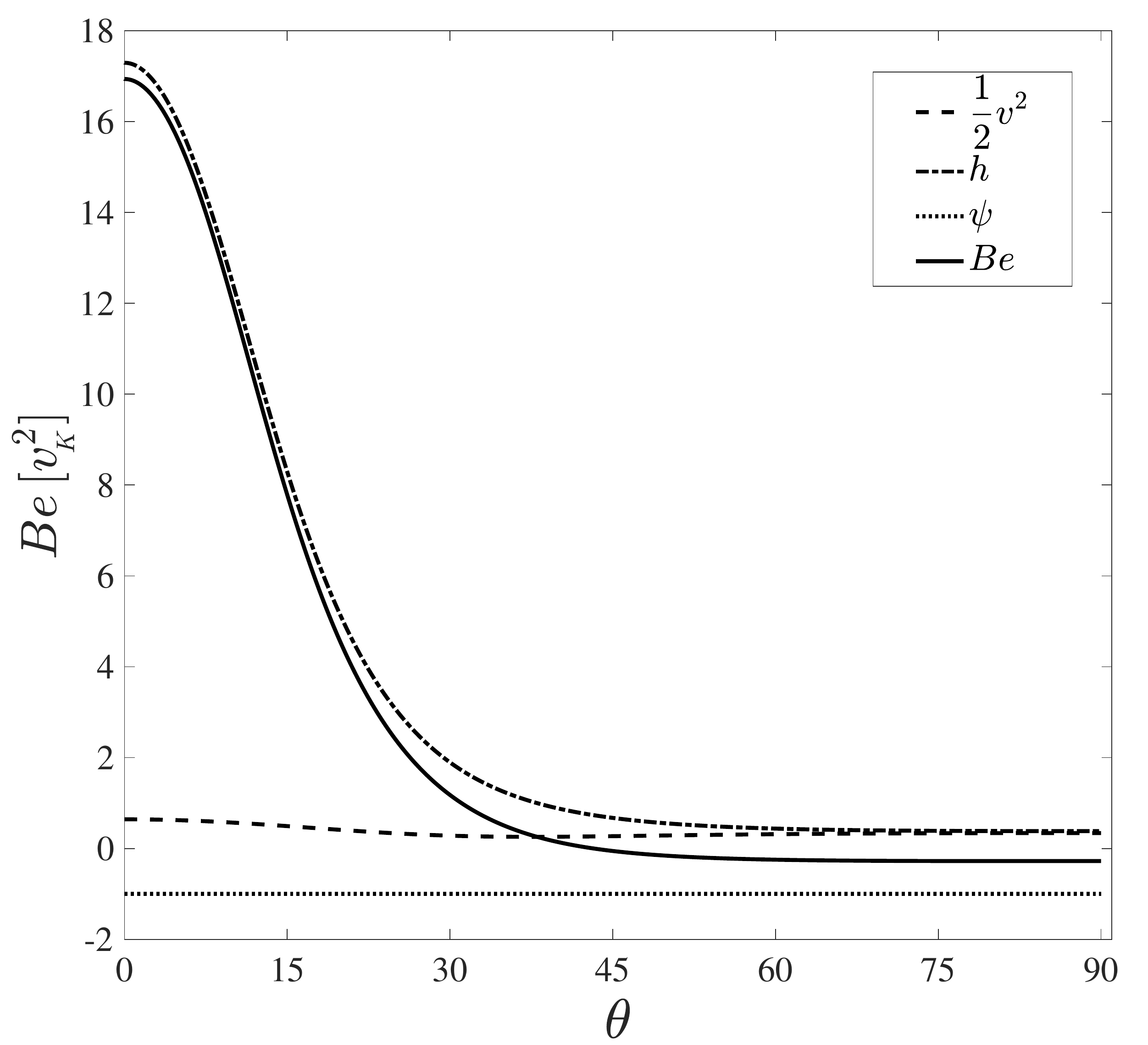}
\centering
\caption{Latitudinal profile of the Bernoulli parameter (solid line),
corresponding kinetic energy, $ \frac{1}{2}\bm{v}^2 $ (dashed line), enthalpy, $ h $ (dash-dotted line),
and gravitational energy, $ \phi $ (dotted line).
\label{bernoulli_parameter}}
\end{figure}

Figure \ref{bernoulli_parameter} shows the Bernoulli parameter
(solid line) and its three terms in Equation (\ref{Bernoulli_parameter}), including kinetic energy
(dashed line), enthalpy (dash-dotted line), and gravitational energy (dotted line). 
From this Figure, the Bernoulli parameter is negative around 
the equatorial plane, and its value becomes larger and positive 
at the high latitudes. In the region near the equatorial plane, 
the gravitational energy is the dominant one and it is followed by the Bernoulli parameter. 
Additionally, in high latitudes, the Bernoulli parameter follows the enthalpy due
to this fact that the density drops faster than the pressure from equatorial plane to
the rotation axis. Therefore, the square of the sound speed, and equivalently the enthalpy 
rises as $ \theta $ angle decreases. 
The trend of the Bernoulli parameter is similar to accretion outflow solution 
in XC97. However, in that solution the Bernoulli parameter 
is always positive independent of $ \theta $.

\subsection{Convective stability}\label{convective}
In this subsection, we investigate the convective stability of hot accretion flows based on 
our self-similar solutions. In this regard, we use the well-known Solberg-H\o iland criterions 
in cylindrical coordinates $ (R, \phi, z) $. If the disc is convectively stable, the two following Solberg-H\o iland criterions must be positive as,

\begin{equation} \label{Hoiland_1}
	\frac{1}{R^{3}} \frac{\partial l^{2}}{\partial R} - \frac{1}{C_{P} \rho} 
	\bm{\nabla} P \cdot \bm{\nabla} S > 0,
\end{equation}

\begin{equation} \label{Hoiland_2}
	- \frac{\partial P}{\partial z} \left( \frac{\partial l^{2}}{\partial R} 
	\frac{\partial S}{\partial z} -  \frac{\partial l^{2}}{\partial z} \frac{\partial S}{\partial R} \right) > 0,
\end{equation}
where $ l \left[= r \sin \theta v_{\phi} \right] $ is the specific angular momentum per unit mass, 
$ C_{P} $ is the specific heat at constant pressure, $ P $ is the total pressure which
equals to the gas pressure in current study, and $ S $ is the entropy defined as,

\begin{equation} \label{entropy}
	dS \propto d \, \ln \left(\frac{P}{\rho^{\gamma}} \right).
\end{equation}
The first criterion can be reduced as,

\begin{equation} \label{Hiland_1_simplified}
	N_\mathrm{eff} = \kappa^{2} + N_{\!_{R}}^{2} + N_{z}^{2} > 0,
\end{equation}
with
\begin{equation} \label{kappa}
	\kappa^{2} = \frac{1}{R^{3}} \frac{\partial l^{2}}{\partial R}, 
\end{equation}
\begin{equation} \label{N2_R}
	N_{\!_{R}}^{2} = - \frac{1}{\gamma \rho} \frac{\partial P}{\partial R} 
	\frac{\partial}{\partial R} \ln \left( \frac{P}{\rho^{\gamma}} \right),
\end{equation}
\begin{equation} \label{N2_z}
	N^{2}_{z} = - \frac{1}{\gamma \rho} \frac{\partial P}{\partial z} 
	\frac{\partial}{\partial z} \ln \left( \frac{P}{\rho^{\gamma}} \right).
\end{equation}

In the above equations, $ N_\mathrm{eff} $, $ \kappa $, $ N_{\!_{R}}^{2} $ and $ N^{2}_{z} $ 
are the effective frequency, the epicyclic frequency,
and the $ R $ and $ z $ components of the Brunt-V\"{a}is\"{a}l\"{a} 
frequency, respectively. The $ \partial P / \partial z $ is always negative, 
therefore the second Solberg-H\o iland criterion can be reduced as,

\begin{equation} \label{Hiland_2_simplified}
	\Delta_{lS} \equiv \frac{\partial l^{2}}{\partial R} \frac{\partial}{\partial z} \ln \left( \frac{P}{\rho^{\gamma}} \right)  -  \frac{\partial l^{2}}{\partial z} \frac{\partial}{\partial R}  \ln \left( \frac{P}{\rho^{\gamma}} \right)  > 0.
\end{equation}
\\
The following transformations will be adopted to find the angular dependency of 
two Solberg-H\o iland criterions in spherical coordinates,

\begin{equation} \label{transformation_R}
	\frac{\partial}{\partial R} = \sin \theta \frac{\partial}{\partial r} + \frac{\cos \theta}{r} \frac{\partial}{\partial \theta}
\end{equation}
\begin{equation} \label{transformation_z}
	\frac{\partial}{\partial z} = \cos \theta \frac{\partial}{\partial r} - \frac{\sin \theta}{r} \frac{\partial}{\partial \theta}.
\end{equation}

\begin{figure}[ht!]
\includegraphics[width=0.49\textwidth]{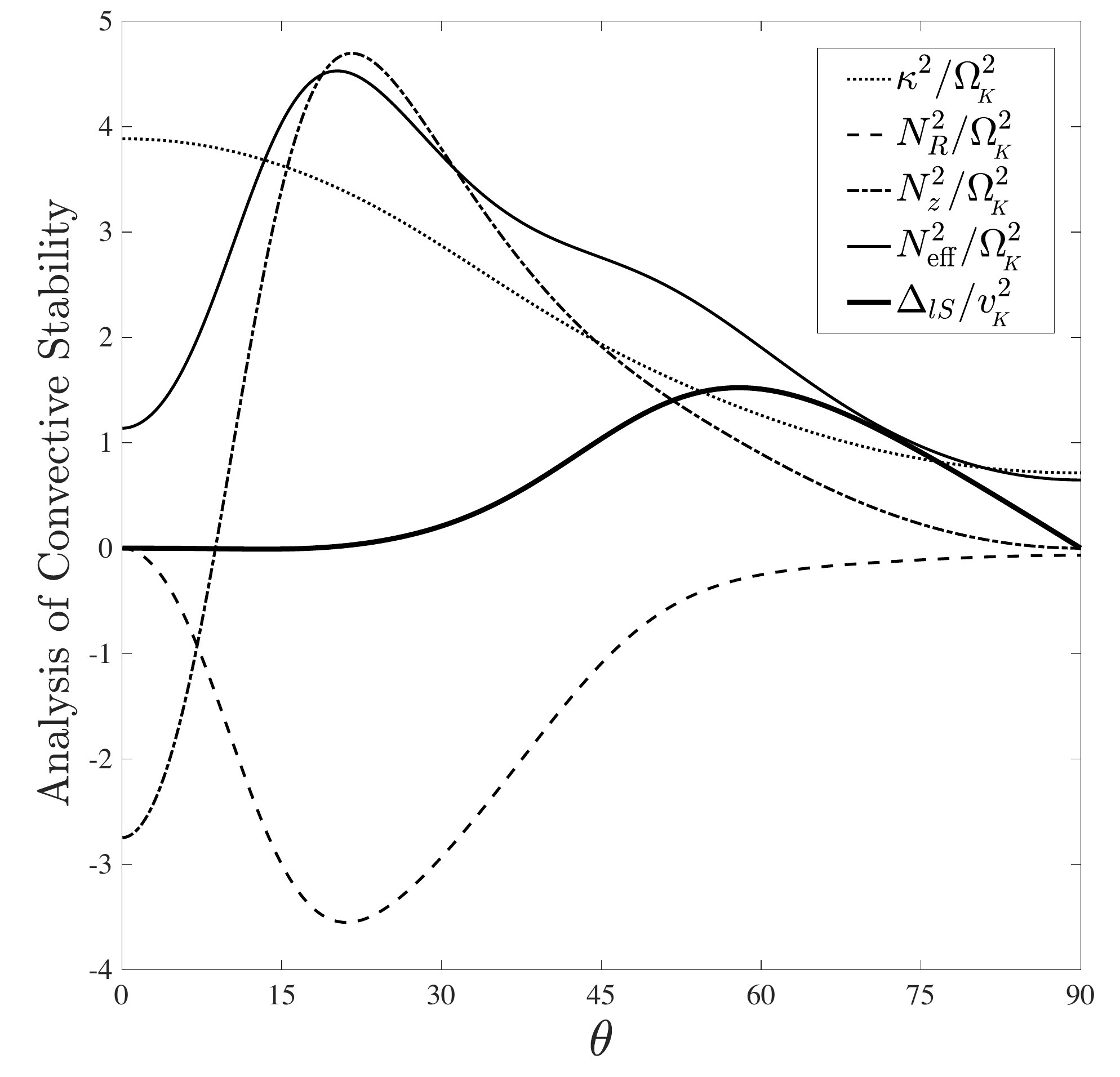}
\caption{Analysis of convective stability. The latitudinal profiles of $ \kappa^2 $, 
$ N_R^2 $, $ N_z^2 $, and $ N_{\mathrm{eff}}^2 $ normalized by $ \Omega_{K}^2 $, 
and $ \Delta_{lS} $ normalized by $ v_K^2 $.
\label{convective}}
\end{figure}

\begin{figure*}[ht!]
\includegraphics[width=\textwidth]{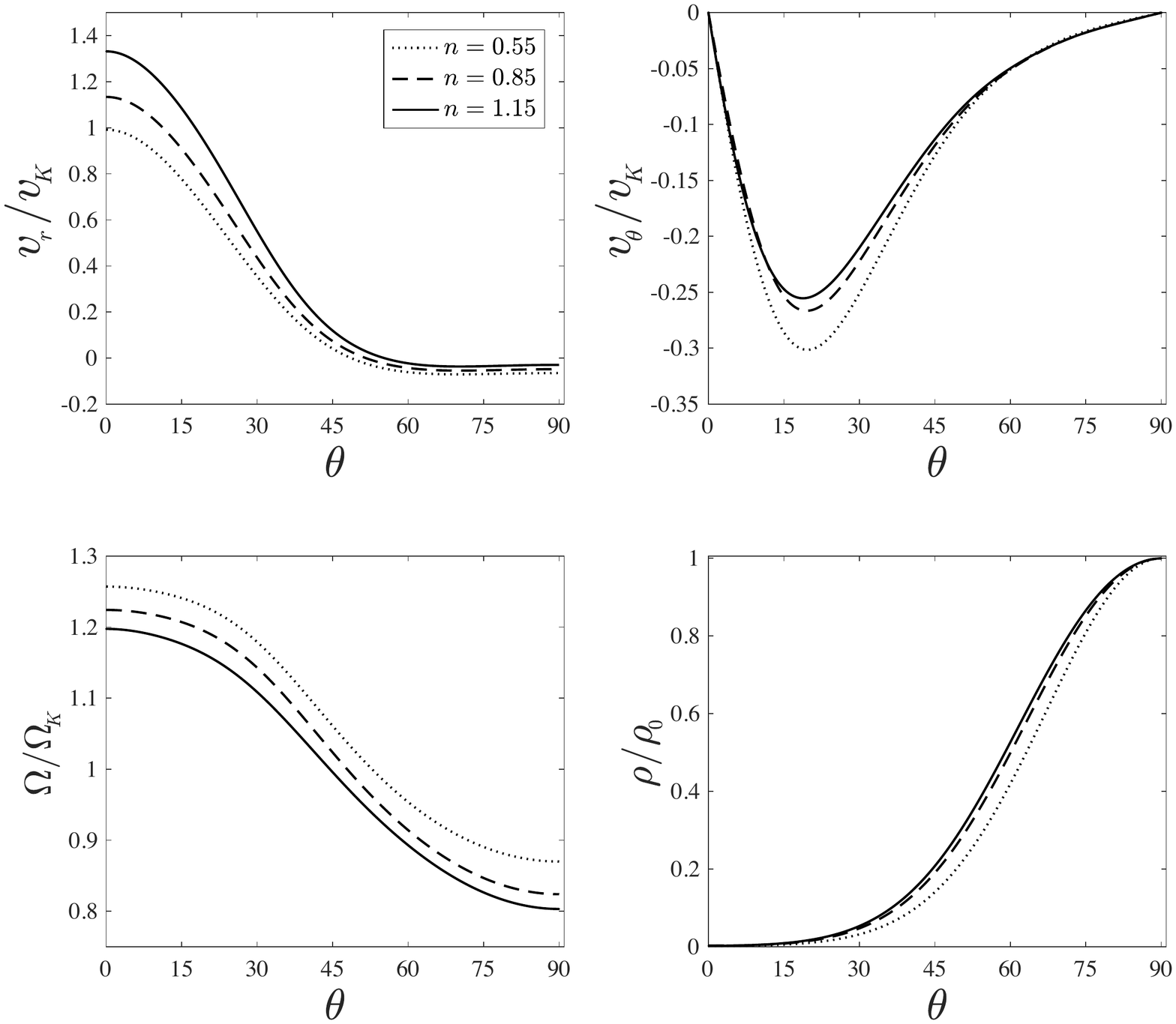} 
\centering
\caption{The dependency of the solution to the density index, $ n $. Here,
$ \alpha = 0.15 $, $ \gamma = 5/3 $ and $ \alpha_{c} = 0.2 $. 
\label{param_n}}
\end{figure*}

In Figure \ref{convective} we show the angular variations of  $ \kappa^{2} $, 
$ N^{2}_{\!_{R}} $, $ N^{2}_{z} $, $ N^{2}_\mathrm{eff} $ normalized by 
$ \Omega^{2}_{K} $ and also $ \Delta_{lS} $ normalized by 
$ v^{2}_{\scriptscriptstyle K} $. We can see that $ \kappa^{2} $ 
is positive while $ N^{2}_{R}$ is negative in whole $ \theta $ angles. 
On the other hand, $ N^{2}_{z} $ is only negative in a small area near the rotationa axis
so, in the rest of the area is positive. Inevitably, $ N^{2}_\mathrm{eff} $ becomes positive
and the first criterion would be satisfied (the light solid line). 
Moreover, as it is shown in this Figure, $ \Delta_{lS} \geq 0 $ in whole domain which means 
the second criterion is also satisfied. Since, both Solberg-H\o iland criterions are satisfied here, 
we conclude that the disc is convectively stable in the presence of thermal conduction. 
Note that, the numerical MHD simulations (\citealt{Narayan et al. 2012}; \citealt{Yuan et al. 2012a}) 
show magnetic field can cause the flow becomes convective stable. We believe that our results
are totally in agreement with the prediction of the numerical MHD simulations. 
We adopted viscosity to mimic the effect of MRI in 
driving angular momentum from the system. In addition,
thermal conduction can convectively stabilize the accretion system in hot accretion mode.
In our future MHD study, we will investigate the convective stability of the hot accretion flow in more details.

\begin{figure*}[ht!]
\includegraphics[width=0.8\textwidth]{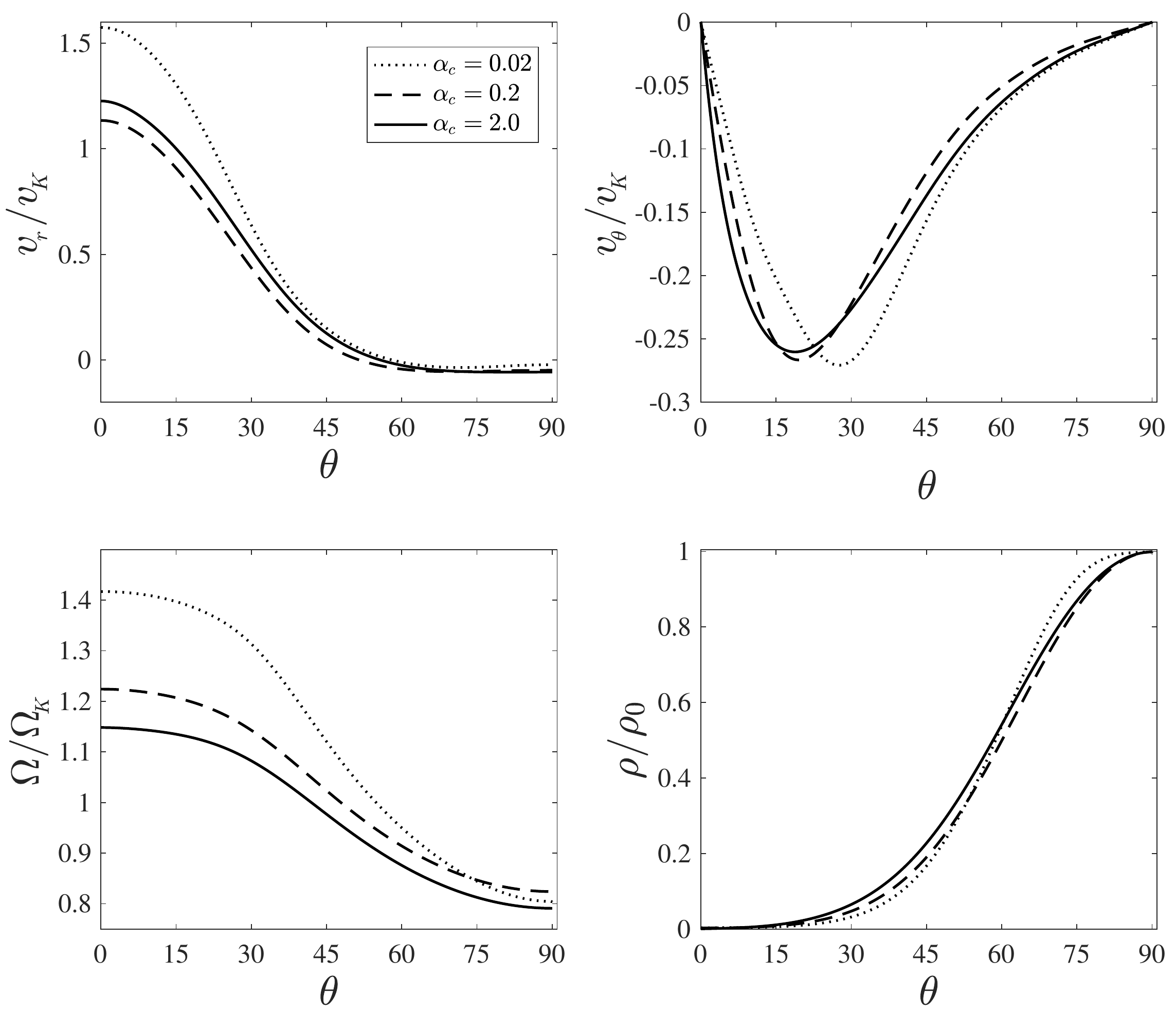} 
\centering
\caption{The dependency of the solution to the conductivity coefficient, $ \alpha_c  $. Here,
$ \alpha = 0.15 $, $ \gamma = 5/3 $ and $ n = 0.85 $.
\label{param_alpha_c}}
\end{figure*}
\begin{figure*}[ht!]
\includegraphics[width=0.8\textwidth]{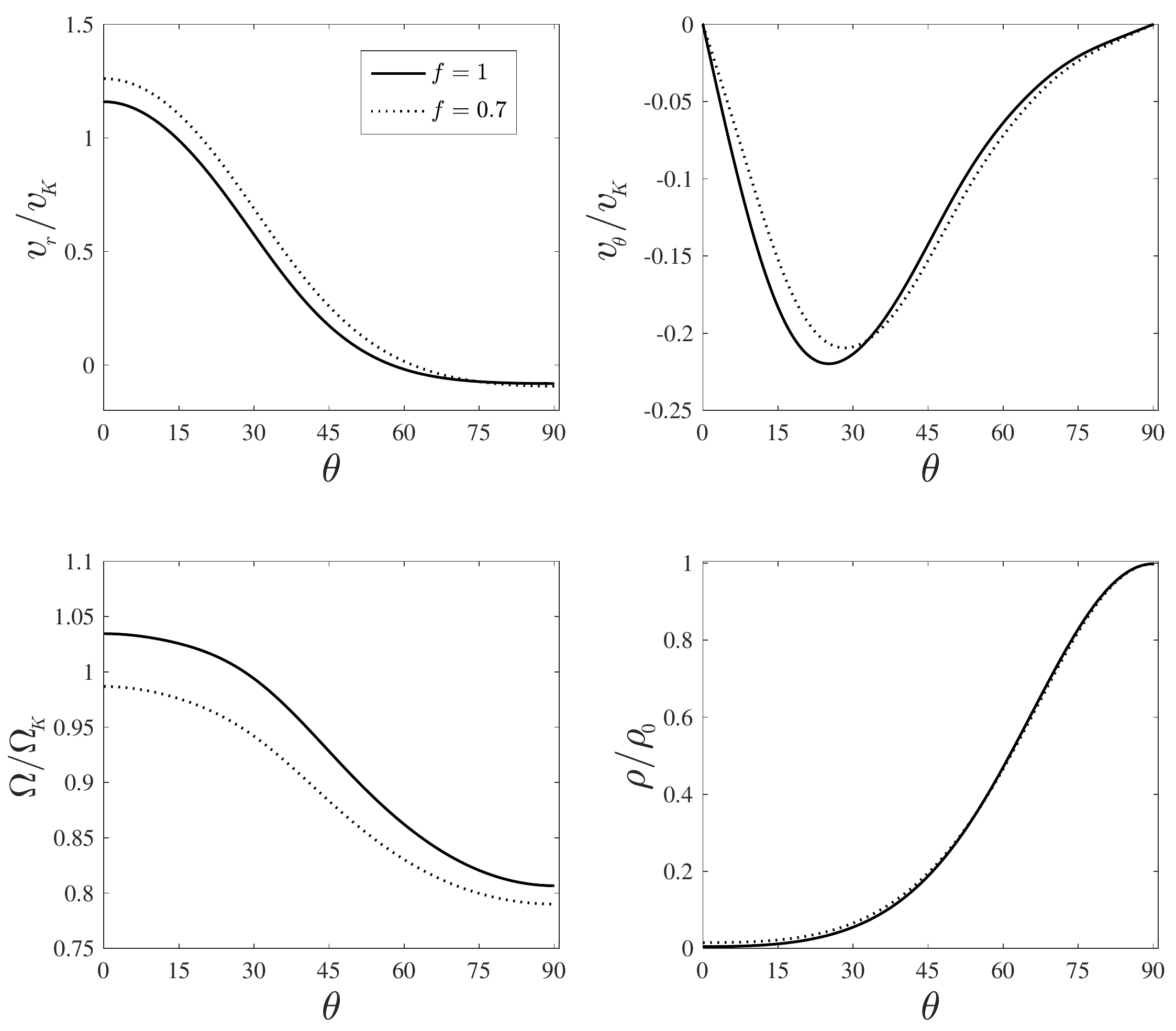} 
\centering
\caption{The dependency of the solution to the advection parameter; $ f = 0.7 $ (dotted line), $ f = 1 $ (solid line). Here,
$ \alpha = 0.2 $, $ \gamma = 5/3 $ and $ n = 0.85 $.
\label{f_param}}
\end{figure*}

\subsection{Dependency of the Solutions on Input Parameters} \label{Dependency_of _the_Solutions}
In this section, we investigate the dependency of the solutions to the 
input parameters including, density index, $ n $, conductivity parameter, 
$ \alpha_c $, and advection parameter, $ f $. As it was mentioned before, similar to XC97 and TM06, 
we obtained the solutions in the whole $ \theta $ direction, from the rotation axis to the equatorial plane, 
showing the strong wind at high latitudes. However, 
in contrast to XC97 which solved the HD equations of hot accretion flow without thermal conduction,
we evinced that the thermal conduction should be inevitably considered in the energy equation (see section \ref{energy_inflow_region}).
Moreover, as introduced in Section \ref{sec:basic_equations}, the density index, $ n $, of the self-similar solutions shows how the density changes along the radius. In the case of $ n = 3/2 $, the solution is similar to TM06 where $ v_{\theta} $ was eliminated from the system of the equations. In section \ref{sec:answer_to_question} , we also showed that for highly non-relativistic cases ($ \gamma=5/3 $) the density index must be $ n < 3/2 $.
\footnote{The density index of the accretion outflow solution of XC97 
is in the same range as we considered, i.e., $ n<3/2 $, where they set $ n = 0.5 $.}
Although, the real accretion systems around the BHs,
where this conclusion may not be satisfied there, happen in the relativistic regime, 
specially when the thermal energy of the gas becomes comparable to (or exceeds) 
the rest mass energy of the electron (\citealt{Chattopadhyay and Ryu 2009}, \citealt{Kumar et al. 2013}).

We have shown the dependency of the results to the index parameter, $ n $, 
in Figure \ref{param_n}. We consider three different values of the density index, i.e., 
$ n = 0.55, 0.85, \, \text{and}\, 1.15 $. From the top left panel of this figure, we can see that the 
maximum amount of the radial velocity of the wind is for $ n = 1.15 $. 
The latitudinal component of the velocity is always negative and becomes null at both boundaries
for all values of $ n $ (see top right panel). In addition, from $ n = 0.55 $ to $ n = 1.15 $, $ v_{\theta} $ 
decreases which shows an opposite behavior in respect to $ v_{r} $. This result can be predictable from
continuity equation, (see equation (\ref{pde_continuity})). In fact, in our current study, this equation 
has two terms and the sum of these terms should be always equal to zero in the whole 
$ \theta $ direction. Therefore, in a fixed density index, any increase in $ v_{r} $ cause a decrease in $ v_{\theta} $. 
The bottom right panel of Figure \ref{param_n} illustrates that the density profile drops faster with
$ n = 0.55 $, rather than $ n = 1.15 $. Also, the flow rotates faster for low density indices. 

The numerical simulations of the hot accretion flow also studied the dependency of the
solutions to the conductivity coefficient, $ \alpha_c $ (e.g., \citealt{Bu et al. 2016}). 
At a fixed radius, from the time averaged of the solutions in steady state, 
they found that the density and pressure change slightly with
increasing $ \alpha_c $. To compare our results with numerical simulations, we also plot 
Figure \ref{param_alpha_c}. We pick three values of the conductivity coefficient 
for this comparison, $ \alpha_c = [0.02, 0.2, 2.0] $.
In the top left panel, it is shown that as the value of $ \alpha_c $ increases, 
the radial velocity at the equator becomes slightly larger 
which is consistent with the results obtained in KS13 (see figure 1 of KS13). 
Moreover, our results show that for the wind region unlike KS13, 
$ v_{r} $ is still rising which is commensurate with the numerical simulations of 
\citealt{Bu et al. 2016}. In the top right panel, $ v_{\theta} $ decreases with increasing 
conductivity coefficient. This trend is consistent with the continuity equation terms discussed above.
However, KS13 found that $ v_{\theta} $ had increasing 
tendency with an enhancement in the thermal conduction (see top panel of figure 2 in KS13). The behavior of $ v_{\theta} $ 
in the present study and also in KS13 are not totally similar since, we impose $ v_{\theta} = 0 $ 
at the rotation axis as a symmetric boundary condition.
From the bottom left panel, we find that the value of $ \alpha_{c} $ affects  
the angular velocity of the flow, where the flow rotates slower with the 
greater values of the conductivity coefficient.
The changes of the angular velocity with thermal conduction is consistent with the result of TM06 (see top right panel of figure 2 in TM06).
From the density profile, the flow indicates a more spherical structure with a rise in thermal conduction
which is consistent with numerical simulations and also KS13 (see top panel of figure 3 in KS13).

We further treated the dependency of the solution to the advection parameter in Figure \ref{f_param}
for two different values; $ f = 0.7 $ (dotted line) and $ f = 1 $ (solid line). 
In top left panel, the radial velocity, $ v_r $, of the wind drops with growing of the advection parameter.
In top right panel, the minimum of the latitudinal velocity, $ v_{\theta} $, moves toward the rotation axis as 
advection parameter increases. Moreover, $ v_{\theta} $ in comparison with $ v_r $ shows inverse behavior
as it increases with the growth of advection parameter. The enhancement of the $ f $ would raise the angular velocity of the gas particles significantly as you can see in the bottom left panel of this figure. However, from the bottom right panel, the density profile  
would not extensively suffer from the changes of the advection parameter\footnote{It should be mentioned here that at small $ f $, the radiation may not be totally ignored as in this paper.}.

In this work, we solved the HD equations with considering thermal conduction 
as well as  the viscosity to mimic the effects of magnetic field. 
This will motivate us to solve the full set of MHD equations to understand the properties and nature of the hot 
accretion flow in a more realistic case. Therefore, in our future studies we mostly focus on
MHD equations and investigative the dependency of the solutions to different configurations
of magnetic field. We will also compare the HD and MHD analytical solutions to find a unique 
and accurate solution for the hot accretion flow which might be useful for numerical simulations.

\section{Summary and Discussion}\label{sec:summary_discussion}

In summary, we have shown thermal conduction is crucial term for 
investigating the inflow-wind structure of hot accretion flows. As argued in 
section \ref{sec:answer_to_question}, thermal conduction is only significant in very low 
accretion rate systems which are suitable for very low luminosity AGNs,
e.g., our Galactic center Sgr $ \mathrm{ A^*} $, super massive black holes in early-type galaxies, and
a considerable number of quiescent X-ray binaries. On this matter, we have solved 
the two-dimensional HD equations of hot accretion flow with the inclusion of the thermal 
conduction and all components of the viscous stress tensor. For simplicity, we adopted 
steady state as well as axisymmetric assumptions and used self-similar approximation
in radial direction. Our integration starts from the 
rotation axis and stops at the equatorial plane so, the solutions will be obtained in the 
full $ r - \theta $ space. We imposed the physical boundary 
conditions at both boundaries and used relaxation method to solve the coupled system 
of equations. 

We have obtained an inflow-wind solution extended
over full range of $ \theta $ direction. The inflow region is around the equatorial plane, 
while the wind region is located at high latitudes around the rotation axis, i.e., 
$ 0  <  \theta \leqslant 52^\circ $, (see Figure \ref{density_temperature}). The density is 
torus-like and its concentration occurs in the equatorial plane and decreases in 
the wind region. From our results, angular velocity behavior shows that wind is able 
to transfer angular momentum outward. Moreover, the gradient of the gas pressure 
is the dominant force in driving wind. Our results also show there is no sonic 
point in our calculation domain. Analysis of the energy balance between advection, thermal conduction and viscous heating 
(Figure \ref{energy}) indicates that while viscous dissipation heats the flow everywhere, 
thermal conduction cools it at the equator and polar region. Therefore, thermal conduction 
conducts the heat from inner regions to outer regions and proceeds as a mechanism for launching 
thermal wind. Moreover, to balance the two terms in the right hand side of the energy equation 
(viscous heating and thermal conduction), advection cools the gas in the disc region while 
in the polar region, acts as a heating mechanism and heats the flow.
The Bernoulli parameter is positive in wind region and negative in the inflow region.
Therefore, it could be still an useful value, if not be an arguable factor, to show the existence 
of wind in the hot accretion flows. We compared our results with previous 
related analytical studies on hot accretion flow with and without thermal conduction.

We also treated the convective stability of the hot accretion flow
in HD case and found that the disc is convectively stable in the presence of the thermal conduction. 
From parameter study of the density index, $ n $, the radial velocity, $ v_{r} $,
increases with growth of $ n $ in the wind region. Whilst, the latitudinal velocity, $ v_{\theta} $, and angular velocity, $ \Omega $,
decrease with increasing of $ n $. In this work, we explored the role of thermal conduction
in hot accretion flows in HD case with parameter study (Figure \ref{param_alpha_c}). 
From our results, thermal conduction did not have effective role to change $ v_{r} $ and $ v_{\theta} $ in inflow region. 
However, in wind region, an inverse behavior were shown
for these two components of the velocity. With an increase in the conductivity coefficient, the radial velocity enhances 
while the latitudinal component drops. Moreover, the disc rotates slower with growth of the thermal conduction strength. 
Thermal conduction also slightly does enlarge the density in the entire of the flow. These results are fully consistent with the numerical simulations of the hot accretion flow (e.g.,  \citealt{Bu et al. 2016}).
To tackle the influence of the advection parameter on the physical variables,
we did a comparison between two values, $ f = 0.7 $ and $ f = 1 $. A rise in the 
advection parameter would increase both the latitudinal and the angular velocities while decrease 
the radial velocity. More, growth of the advection parameter could not make substantial change 
in the density profile of the accretion flow.

There are several caveats in this study which will be improved in our future studies. 
The first one is that we only solved the HD equations of hot accretion flow. In a real 
accretion flow, angular momentum is transferred
by Maxwell stress associated with MHD turbulence driven by MRI. 
In addition, magnetic field is one of the driving mechanisms for wind production. 
With this aim, we will solve the MHD equations of the accretion flow. In our next papers we mainly 
focus on different magnetic field configurations on the structure of the hot accretion flow 
and compare the results with HD case. A further simplification here is that we adopted 
a single one temperature fluid. In hot accretion model, it is expected the ions 
be much more hotter than the electrons (see, \citealt{Rees et al. 1982}; \citealt{Yuan and Narayan 2014}). 
Thus, two different energy equations for electrons and ions should be solved. 

\section*{Aknowledgments}
The authors would like to thank the referee for his/her thoughtful and constructive comments.
Fatemeh Zahra Zeraatgari is supported by the National Natural Science Foundation of China (grant No. 12003021).
This work is supported by the Science Challenge Project of China (grant No. TZ2016002), 
the China Postdoctoral Science Foundation (grants No. 2019M663665, 2020M673371).
De-Fu Bu is supported by the Natural Science Foundation of China (grant No. 11773053).
Amin Mosallanezhad also thanks the support of Dr. X. D. Zhang at the Network Information Center of Xi'an Jiaotong University.
The computation has made use of the High Performance Computing (HPC) platform of Xi'an Jiaotong University.

\appendix

\section{Energy Equation presented in Xu \& Chen 1997} \label{enegy_xu_chen_appendix}

XC97 solved the HD equations of hot accretion flows which was very similar to the equations 
described in this paper. The only difference is the definition of the energy equations. 
The energy equation described in XC97 can be written as,

\begin{equation}  \label{energy_XC97}
	Q_\mathrm{adv} = f Q_\mathrm{vis},
\end{equation}
with

\begin{equation} \label{energy_XC97_Q_adv}
	Q_\mathrm{adv}  = \rho \frac{\mathrm{d} e}{\mathrm{d} t} - \frac{p}{\rho} \frac{\mathrm{d} \rho}{\mathrm{d} t}, 
\end{equation}

\begin{equation} \label{Q_vis}
	Q_\mathrm{vis} =  \nabla \bm{v} : \bm{\sigma},
\end{equation}

By imposing steady state, $ \partial / \partial t = 0 $, and axisymmetric, 
$ \partial/ \partial \phi = 0 $, assumptions, the components of stress 
tensor, $ \sigma $, in the spherical coordinates are given by:

\begin{equation} \label{sigma_rr}
	\sigma_{rr} = \rho \nu  \left[ 2 \frac{\partial v_{r}}{\partial r} - \frac{2}{3} \left( \nabla \cdot \bm{v} \right) \right],
\end{equation}

\begin{equation} \label{sigma_tt}
	\sigma_{\theta \theta} =  \rho \nu \left[ 2 \left( \frac{1}{r} \frac{\partial v_{\theta}}{\partial \theta} + \frac{v_{r}}{r}  
	\right) - \frac{2}{3} \left( \nabla \cdot \bm{v}  \right) \right], 
\end{equation}

\begin{equation} \label{sigma_pp}
	\sigma_{\phi \phi} = \rho \nu \left[ 2 \left( \frac{v_{r}}{r} +  \frac{v_{\theta} \cot \theta}{r}  
	\right) - \frac{2}{3} \left( \nabla \cdot \bm{v}  \right) \right], 
\end{equation}

\begin{equation} \label{sigma_rt}
	\sigma_{r \theta} = \rho \nu  \left[ r \frac{\partial}{\partial r} \left( \frac{v_{\theta}}{r} \right) + \frac{1}{r} \frac{\partial v_{r}}{\partial \theta} 
	 \right], 
\end{equation}

\begin{equation} \label{sigma_rp}
	\sigma_{r \phi} = \rho \nu \left[ r \frac{\partial}{\partial r} \left(\frac{v_{\phi}}{r} \right) + \frac{1}{r \sin \theta} \frac{\partial v_{r}}{\partial \phi} \right],
\end{equation}

\begin{equation} \label{sigma_tp}
	\sigma_{\theta \phi} = \rho \nu \left[ \frac{\sin \theta}{r}  \frac{\partial}{\partial \theta}  
	\left( \frac{v_{\phi}}{\sin \theta} \right) + \frac{1}{r \sin \theta} \frac{\partial v_{\theta}}{\partial \phi} \right],
\end{equation}

where $ \nabla \cdot \bm{v} $ is written as
\begin{equation} \label{deldotv}
	\nabla \cdot \bm{v} = \frac{1}{r^{2}} \frac{\partial}{\partial r} \left( r^{2} v_{r} \right) + \frac{1}{r \sin \theta} \frac{\partial}{\partial \theta} \left( v_{\theta} \sin \theta \right).
\end{equation}
 
Substituting the above terms into the advection and the viscous terms of the energy equations, we have,
\begin{equation} \label{Q_adv_XC97}
	Q_\mathrm{adv} = \rho \left[ v_{r} \frac{\partial e}{\partial r}   + \frac{v_{\theta}}{r}  \frac{\partial e}{\partial \theta} \right] - \frac{p}{\rho} \left[ v_{r} \frac{\partial \rho}{\partial r}   + \frac{v_{\theta}}{r}  \frac{\partial \rho}{\partial \theta} \right] 
\end{equation}

\begin{multline} \label{Q_vis_XC97}
	Q_\mathrm{vis} =  f \Bigg[ \frac{\partial v_{r}}{\partial r} \sigma_{rr} + \frac{\partial v_{\theta}}{\partial r} \sigma_{r \theta} + \frac{\partial v_{\phi}}{\partial r} \sigma_{r \phi} + \frac{1}{r} \left( \frac{\partial v_{r}}{\partial \theta} - v_{\theta} \right) \sigma_{r \theta} + \frac{1}{r} \left( \frac{\partial v_{\theta}}{\partial \theta} + v_{r} \right) \sigma_{\theta \theta} + \frac{1}{r} \frac{\partial v_{\phi}}{\partial \theta} \sigma_{\theta \phi} \\
	 - \frac{v_{\phi}}{r} \left( \sigma_{r \phi} + \sigma_{\theta \phi} \cot \theta \right) + \frac{\sigma_{\phi \phi}}{r} \left( v_{r} + v_{\theta} \cot \theta \right) \Bigg]
\end{multline}

They assumed fully advection case, i.e., $ f = 1 $. Substituting self-similar
approximation, equation (\ref{rho_selfsimilar})-(\ref{p_gas_selfsimilar}), into the above terms, 
the radial dependency will be removed. Therefore, the dimensionless form of 
the energy equation described in XC97 can be written as,

\begin{equation} \label{q_adv_vis}
	q_\mathrm{adv} = q_\mathrm{vis}
\end{equation}
with

\begin{equation} \label{q_adv_XC97}
	q_\mathrm{adv} = \left[ n  - \frac{1}{ \gamma - 1} \right] p_\mathrm{g} v_{r} + \frac{v_{\theta}}{\gamma - 1}  \left[ \frac{\mathrm{d} p_\mathrm{g}}{\mathrm{d}\theta} - \gamma \frac{p_\mathrm{g}}{\rho} \frac{\mathrm{d} \rho}{\mathrm{d} \theta} \right],
\end{equation}

\begin{multline} \label{q_vis_XC97}
	q_\mathrm{vis} = \alpha p_\mathrm{g} \Biggr[ \frac{1}{2} v_{r}^{2} + 2 \left( v_{r} + \frac{\mathrm{d} v_{\theta}}{\mathrm{d} \theta} \right)^2 + 2  \left( v_{r} + v_{\theta} \cot \theta \right)^{2} + \frac{1}{4} \left( 2 \frac{\mathrm{d} v_{r}}{\mathrm{d} \theta} - 3 v_{\theta} \right)^{2} + \left[ \frac{9}{4} \Omega^{2} + \left( \frac{\mathrm{d}\Omega}{\mathrm{d}\theta} \right)^{2} \right] \sin^{2} \theta \\
	- \frac{2}{3} \left( \frac{3}{2} v_{r} + \frac{\mathrm{d}v_{\theta}}{\mathrm{d} \theta} + v_{\theta} \cot \theta \right)^2 \Biggr].	
\end{multline}

\section{Partial Differential Equations in Spherical Coordinates} \label{pde_appendix}

In this study to reach the final PDEs governing the system, 
we adopt spherical coordinates $ (r,\theta,\phi) $. We consider the disc to be axisymmetric and 
steady state. We further assume the velocity field consists of all its components as 
$ \bm{v} = v_{r} \hat{r} + v_{\theta} \hat{\theta} + v_{\phi} \hat{\phi} $.
All components of the viscous stress tensor will be included as described in equations 
(\ref{sigma_rr})-(\ref{sigma_tp}). By imposing the assumptions and definitions introduced 
in section \ref{sec:basic_equations}, the equations (\ref{eq:continuity})-(\ref{eq:energy}) 
will be reduced to the following system of PDEs. Thus, we can rewrite the continuity equation as,

\begin{equation}\label{pde_continuity}
	\frac{1}{r^{2}} \frac{\partial}{\partial r} \left( r^{2} \rho v_{r} \right) + \frac{1}{r \sin \theta} \frac{\partial}{\partial \theta} \left( \rho v_{\theta} \sin\theta \right) = 0.
\end{equation}
\\
Three components of the equation of motion, equation (\ref{eq:momentum}), can be read as,

\begin{equation}\label{pde_momentun1}
  \rho \left[ v_{r} \frac{\partial v_{r}}{\partial r}  + \frac{v_{\theta}}{r} \left( \frac{\partial v_{r}}{\partial \theta} - v_{\theta} \right) - \frac{v_{\phi}^{2}}{r} \right] = - \frac{GM \rho}{r^{2}} - \frac{\partial p}{\partial r}  + \frac{1}{r^{2}} \frac{\partial}{\partial r} \left( r^{2} \sigma_{rr} \right) + \frac{1}{r \sin \theta} \frac{\partial}{\partial \theta} \left( \sin \theta \sigma_{r \theta} \right) - \frac{1}{r} \left( \sigma_{\theta \theta} + \sigma_{\phi \phi} \right),
\end{equation}

\begin{equation}\label{pde_momentum2}
  \rho \left[  v_{r} \frac{\partial v_{\theta}}{\partial r} + \frac{v_{\theta}}{r} \left( \frac{\partial v_{\theta}}{\partial \theta} + v_{r} \right) - \frac{v_{\phi}^{2}}{r} \cot \theta \right] = - \frac{1}{r} \frac{\partial p}{\partial \theta} + \frac{1}{r^{2}} \frac{\partial}{\partial r} \left( r^{2} \sigma_{r \theta} \right) + \frac{1}{r \sin \theta} \frac{\partial}{\partial \theta} \left( \sin \theta \sigma_{\theta \theta} \right) + \frac{1}{r} \left( \sigma_{r \theta} - \sigma_{\phi \phi} \cot \theta \right),
\end{equation}

\begin{equation}\label{pde_momentum3}
  \rho \left[ v_{r} \frac{\partial v_{\phi}}{\partial r} + \frac{v_{\theta}}{r} \frac{\partial v_{\phi}}{\partial \theta} + \frac{v_{\phi}}{r} \left( v_{r} +  v_{\theta} \cot\theta \right)  \right] =   \frac{1}{r^{2}} \frac{\partial}{\partial r} \left( r^{2} \sigma_{r \phi} \right) + \frac{1}{r \sin \theta} \frac{\partial}{\partial \theta} \left( \sin \theta \sigma_{\theta \phi} \right) + \frac{1}{r} \left( \sigma_{r \phi} + \sigma_{\theta \phi} \cot \theta \right),
\end{equation}
\\
And finally, the energy equation can be written as,

\begin{multline} \label{pde_energy}
		\rho \left[ v_{r} \frac{\partial e}{\partial r}   + \frac{v_{\theta}}{r}  \frac{\partial e}{\partial \theta} \right] - \frac{p}{\rho} \left[ v_{r} \frac{\partial \rho}{\partial r}   + \frac{v_{\theta}}{r}  \frac{\partial \rho}{\partial \theta} \right]  = f \Bigg[ \frac{\partial v_{r}}{\partial r} \sigma_{rr} + \frac{\partial v_{\theta}}{\partial r} \sigma_{r \theta} + \frac{\partial v_{\phi}}{\partial r} \sigma_{r \phi} + \frac{1}{r} \left( \frac{\partial v_{r}}{\partial \theta} - v_{\theta} \right) \sigma_{r \theta} + \frac{1}{r} \left( \frac{\partial v_{\theta}}{\partial \theta} + v_{r} \right) \sigma_{\theta \theta} \\
		+ \frac{1}{r} \frac{\partial v_{\phi}}{\partial \theta} \sigma_{\theta \phi}  - \frac{v_{\phi}}{r} \left( \sigma_{r \phi} + \sigma_{\theta \phi} \cot \theta \right) + \frac{\sigma_{\phi \phi}}{r} \left( v_{r} + v_{\theta} \cot \theta \right) \Bigg] + \frac{1}{r^2} \frac{\partial}{\partial r}\left( r^2 \chi \frac{\partial T}{\partial r} \right) + \frac{1}{r \sin\theta}\frac{\partial}{\partial \theta} \left( \frac{\chi \sin\theta}{r} \frac{\partial T}{\partial \theta} \right).
\end{multline}
\\
\\  
\section{Ordinary Differential Equations} \label{ode_appendix}

By substituting self-similar solutions presented in section \ref{sec:basic_equations} 
into the partial differential equations (\ref{pde_continuity})–(\ref{pde_energy}), 
the following coupled ODEs in the $ \theta $ direction will be obtained
\footnote{Note that, for simplicity, in equations (\ref{ode_continuity})-(\ref{ode_energy}), 
we remove the $ \theta $ dependency of the variables as well as their derivatives.}:

\begin{equation} \label{ode_continuity}
	\left[ \left(\frac{3}{2} - n\right) \bar{v}_{r} + \frac{\mathrm{d}\bar{v}_{\theta}}{\mathrm{d}\theta} + \bar{v}_{\theta} \cot \theta \right] \bar{\rho} + \bar{v}_{\theta} \frac{\mathrm{d} \bar{\rho}}{\mathrm{d}\theta} = 0
\end{equation}

\begin{multline} \label{ode_mom1}
	\bar{\rho} \left[ - \frac{1}{2} \bar{v}_r^2 + \bar{v}_{\theta}\frac{\mathrm{d}\bar{v}_r}{\mathrm{d}\theta} - \bar{v}_{\theta}^2 - \bar{\Omega}^2 \sin^2 \theta \right]
	= - \bar{\rho} + \left( n + 1 \right) \bar{p}_\mathrm{g} + \alpha \left( \frac{\mathrm{d} \bar{v}_{r}}{\mathrm{d} \theta} - \frac{3}{2} \bar{v}_{\theta}  \right) \frac{\mathrm{d} \bar{p}_\mathrm{g}}{\mathrm{d} \theta}	 \\
	+ \alpha  \bar{p}_\mathrm{g} \left[  \frac{2}{3} \left( n + 1 \right) \left( 3 \bar{v}_{r} + \bar{v}_{\theta} \cot \theta + \frac{\mathrm{d} \bar{v}_{\theta}}{\mathrm{d} \theta} \right) - \frac{7}{2} \left( \frac{\mathrm{d} \bar{v}_{\theta}}{\mathrm{d} \theta} + \bar{v}_{\theta} \cot \theta \right) + \frac{\mathrm{d}^{2} \bar{v}_{r}}{\mathrm{d} \theta^{2}}  + \frac{\mathrm{d} \bar{v}_{r}}{\mathrm{d} \theta} \cot \theta - 6 \bar{v}_{r} \right], 
\end{multline}

\begin{multline} \label{ode_mom2}
	\bar{\rho} \left[ \frac{1}{2} \bar{v}_r \bar{v}_{\theta} + \bar{v}_{\theta} \frac{\mathrm{d}\bar{v}_{\theta}}{\mathrm{d}\theta} -\bar{\Omega}^2 \sin \theta \cos \theta \right] = 
	- \frac{\mathrm{d} \bar{p}_\mathrm{g}}{\mathrm{d} \theta} + \alpha \left( \bar{v}_{r} + \frac{4}{3} \frac{\mathrm{d} \bar{v}_{\theta}}{\mathrm{d} \theta} - \frac{2}{3} \bar{v}_{\theta} \cot \theta \right) \frac{\mathrm{d} \bar{p}_\mathrm{g}}{\mathrm{d} \theta} \\
	+ \alpha \bar{p}_\mathrm{g} \left[  \left( n + 1 \right) \left( \frac{3}{2} \bar{v}_{\theta} - \frac{\mathrm{d} \bar{v}_{r}}{\mathrm{d} \theta} \right) + 4 \frac{\mathrm{d} \bar{v}_{r}}{\mathrm{d} \theta} + \frac{4}{3} \left( \frac{\mathrm{d}^{2} \bar{v}_{\theta}}{\mathrm{d} \theta^{2}} + \frac{\mathrm{d} \bar{v}_{\theta}}{\mathrm{d} \theta} \cot \theta - \bar{v}_{\theta} \csc^{2} \theta \right) - \frac{5}{2} \bar{v}_{\theta}  \right], 
\end{multline}

\begin{equation}\label{ode_mom3}
\bar{\rho} \left[ \frac{1}{2} \bar{v}_r \bar{\Omega} + 2 \bar{v}_{\theta} \bar{\Omega} \cot \theta + \bar{v}_{\theta} \frac{\mathrm{d}\bar{\Omega}}{\mathrm{d}\theta} \right] =
\alpha \bar{p}_\mathrm{g} \left[ \frac{3}{2} \left(n - 2 \right) \bar{\Omega} +  3 \frac{\mathrm{d}\bar{\Omega}}{\mathrm{d} \theta} \cot \theta + \frac{\mathrm{d}^{2} \bar{\Omega}} {\mathrm{d} \theta^{2}} \right] + \alpha \frac{\mathrm{d} \bar{\Omega}}{\mathrm{d} \theta} \frac{\mathrm{d} \bar{p}_\mathrm{g}}{\mathrm{d} \theta},
\end{equation}

\begin{multline}\label{ode_energy}
\left( n  - \frac{1}{ \gamma - 1} \right) \bar{p}_\mathrm{g} \bar{v}_{r} + \frac{\bar{v}_{\theta}}{\gamma - 1}  \left( \frac{\mathrm{d} \bar{p}_\mathrm{g}}{\mathrm{d}\theta} - \gamma \frac{\bar{p}_\mathrm{g}}{\bar{\rho}} \frac{\mathrm{d} \bar{\rho}}{\mathrm{d} \theta} \right) = \alpha f \bar{p}_\mathrm{g} \Biggr[ \frac{1}{2} \bar{v}_{r}^{2} + 2 \left( \bar{v}_{r} + \frac{\mathrm{d} \bar{v}_{\theta}}{\mathrm{d} \theta} \right)^2 + 2  \left( \bar{v}_{r} + \bar{v}_{\theta} \cot \theta \right)^{2} \\
+ \frac{1}{4} \left( 2 \frac{\mathrm{d} \bar{v}_{r}}{\mathrm{d} \theta} - 3 \bar{v}_{\theta} \right)^{2} + \left[ \frac{9}{4} \bar{\Omega}^{2}
+ \left( \frac{\mathrm{d}\bar{\Omega}}{\mathrm{d}\theta} \right)^{2} \right] \sin^{2} \theta - \frac{2}{3} \left( \frac{3}{2} \bar{v}_{r} + \frac{\mathrm{d}\bar{v}_{\theta}}{\mathrm{d} \theta} + \bar{v}_{\theta} \cot \theta \right)^2 \Biggr] \\
+ \alpha_{c}  \left[ \left( n - \frac{1}{2} \right) \bar{p}_\mathrm{g} + \left( \frac{\mathrm{d} \bar{p}_\mathrm{g}}{\mathrm{d} \theta} - \frac{\bar{p}_\mathrm{g}}{\bar{\rho}} \frac{\mathrm{d} \bar{\rho}}{\mathrm{d} \theta} \right) \cot \theta + \frac{\mathrm{d}^{2} \bar{p}_\mathrm{g}}{\mathrm{d} \theta^{2} } - \frac{1}{\bar{\rho}} \frac{\mathrm{d} \bar{\rho}}{\mathrm{d} \theta} \frac{\mathrm{d} \bar{p}_\mathrm{g}}{\mathrm{d} \theta}  + \frac{\bar{p}_\mathrm{g}}{\bar{\rho}^{2}} \left( \frac{\mathrm{d} \bar{\rho}}{\mathrm{d} \theta} \right)^{2} - \frac{\bar{p}_\mathrm{g}}{\bar{\rho}} \frac{\mathrm{d}^{2} \bar{\rho}}{\mathrm{d} \theta^{2}} \right].
\end{multline}


\bibliography{sample63}{}
\bibliographystyle{aasjournal}



\end{document}